%% file: main.tex
\documentclass[%
 aip,
 amsmath,amssymb,
 reprint,%
]{revtex4-1}

\usepackage{graphicx}% Include figure files
\usepackage{dcolumn}% Align table columns on decimal point
\usepackage{bm}% bold math
\usepackage[normalem]{ulem}
\usepackage{xcolor}

\usepackage[utf8]{inputenc}
\usepackage[T1]{fontenc}
\usepackage{mathptmx}
\usepackage{etoolbox}

\usepackage{subcaption}
\usepackage{multirow, booktabs, threeparttable} %added by LD
\usepackage{todonotes} %Added by FZ (notes)
\usepackage{soul} % Para resaltado (highlighting)
\makeatletter
\def\selectlanguage#1{\relax} %Added by LD
\def\@email#1#2{%
 \endgroup
 \patchcmd{\titleblock@produce}
  {\frontmatter@RRAPformat}
  {\frontmatter@RRAPformat{\produce@RRAP{*#1\href{mailto:#2}{#2}}}\frontmatter@RRAPformat}
  {}{}
}%
\makeatother
\begin{document}

\preprint{AIP/123-QED}

\title[Numerical Modeling of Flow and Air Entrainment in Hydraulic Jumps]{Numerical Modeling of Flow and Air Entrainment in Hydraulic Jumps for a Wide Range of Froude Numbers}

\author{L. D'Angelo}
 \email{dangelo.leiza@ing.unlp.edu.ar}
 \affiliation{Department of Hydraulics, National University of La Plata, La Plata 1900, Argentina}%Lines break automatically or can be forced with \\

 \author{F. Zabaleta}
 \affiliation{Center for Turbulence Research, Stanford University, Stanford, California 94305, USA}%Lines break automatically or can be forced with \\
 
\author{G.E. Spadari}%
 %\email{gespadari@ucdavis.edu}
\affiliation{Department of Civil and Environmental Engineering, University of California, Davis, California 95616, USA%\\This line break forced with \textbackslash\textbackslash
}%
\author{P. Consoli-Lizzi}
 \affiliation{Department of Hydraulics, National University of La Plata, La Plata 1900, Argentina}%Lines break automatically or can be forced with \\
 
\author{F.A. Bombardelli}
\affiliation{Department of Civil and Environmental Engineering, University of California, Davis, California 95616, USA}%

\date{\today}

\begin{abstract}
The numerical modeling of hydraulic jumps remains challenging due to complex interactions among free-surface deformation, air entrainment and detrainment, and turbulent bubble transport. Whereas accurate prediction of these flows is essential for the design of hydraulic structures, existing high-fidelity tools require prohibitive computational resources for engineering applications. This study implements a three-phase mixture model based on an Unsteady Reynolds-Averaged Navier Stokes (URANS) framework, to numerically simulate flow and air entrainment across twelve hydraulic jumps with Froude numbers ranging from $1.98$ to $8.48$, representing the first systematic analysis for such a comprehensive range of Froude numbers. The model accurately represents time-averaged velocity fields and air concentration profiles, as well as dynamic features including jump toe oscillation and free-surface deformation, showing good agreement with experimental data from seven facilities. Compared to Improved Delayed Detached Eddy Simulations (IDDES), the proposed approach achieves similar accuracy with approximately 400-fold fewer cells and a 300-fold reduction in computational cost. The investigation shows that the selection of turbulence closure affects the accuracy of the prediction of air entrainment. These findings establish the three-phase mixture approach as a practical engineering tool for hydraulic jump simulation, offering an effective balance of accuracy and computational cost.
\end{abstract}

\maketitle

\input{I_Introduction}
\input{II_Theoretical_Model_and_Numerical_Implementation}
\input{III_Numerical_model_setup_and_validation}
\input{IV_Validation_and_parameter_selection}

\input{V_Results}
\input{VI_Conclusions}
\begin{acknowledgments}
The authors gratefully acknowledge the financial support for this research provided by the Hydromechanics Laboratory at the National University of La Plata. In particular, we thank very much its Director, Prof. Sergio O. Liscia. Computational resources were provided by the cluster Pirayu, acquired with funds from the Agencia Santafesina de Ciencia, Tecnología e Innovación (ASACTEI), Government of the Province of Santa Fe, Argentina (Project AC-00010-18, Resolution No.~117/14), which is part of the National High-Performance Computing System (SNCAD) of the Argentine Ministry of Science and Technology; and by UNC Supercómputo (CCAD) -- Universidad Nacional de Córdoba, also part of SNCAD.
\end{acknowledgments}
\newpage
\input{VII_Appendix}

\newpage
\section*{References}
%\bibliography{aipsamp}
%merlin.mbs aipnum4-1.bst 2010-07-25 4.21a (PWD, AO, DPC) hacked
%Control: key (0)
%Control: author (8) initials jnrlst
%Control: editor formatted (1) identically to author
%Control: production of article title (0) allowed
%Control: page (1) range
%Control: year (1) truncated
%Control: production of eprint (0) enabled
%

\end{document}

%% file: I_Introduction.tex
\section{Introduction}
Hydraulic jumps constitute abrupt transitions from supercritical to subcritical flow and are characterized by a rapid increase in water depth, local turbulence production, air entrainment and detrainment, and energy dissipation~\cite{kundu_fluid_2016}. They occur when the channel slope transitions from steep to mild, or take place downstream of sluice gates releasing into supercritical conditions in otherwise subcritical flow condition; further, they can be deliberately generated in hydraulic structures to dissipate energy. These flows therefore play an essential role in stilling basins~\cite{peterka_engineering_1984, hager_energy_1992}. The intense air-water interactions enhances dissolved oxygen levels, improving downstream water quality~\cite{chapman_ambient_1986, geldert_modeling_1998, allan_stream_2007}. However, excessive air entrainment in hydraulic jumps can lead to gas bubble disease in downstream fish populations when total dissolved gas exceeds 110\% saturation~\cite{weitkamp_review_1980, domitrovic_mortandad_1994, politano_multidimensional_2007, epa_quality_1986}.% Multiphase air-water flows in hydraulic structures have long posed significant challenges for both experimental measurements and numerical simulations.

\begin{figure}%[htb]
\includegraphics[width=\columnwidth]{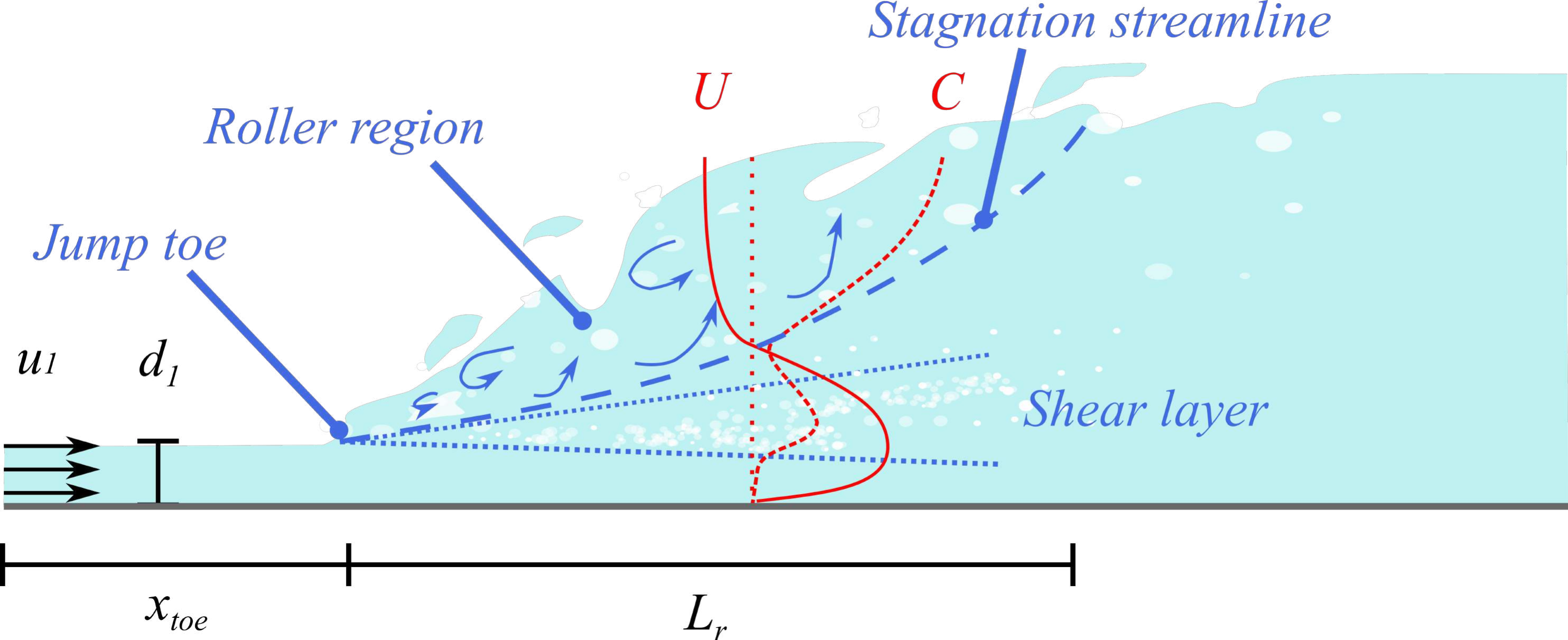}
\caption{\label{fig:sketch} Definition sketch of a free hydraulic jump.}
\end{figure}

%Flow Structure and Morphology
The internal structure of a hydraulic jump exhibits distinct regions with characteristic flow patterns (Figure \ref{fig:sketch}): At the jump toe (impingement point), the incoming supercritical flow generates a turbulent shear layer where large-scale vortical structures interact with entrained air bubbles~\cite{rajaratnam_profile_1968, resch_mesures_1971, chanson_air_1995-1, chanson_experimental_1997, chanson_hydraulic_2011}. Above this layer, a surface roller develops, with recirculating flow and intense free-surface oscillations, extending from the end of the roller (where stagnation streamline reaches the free surface) back to the jump toe~\cite{hager_energy_1992}. Below the shear layer, a developing boundary layer interacts with the channel bed, characterized by steep velocity gradients.

Jump typology depends primarily on the incident Froude number (Fr$_1 = u_1/\sqrt{g d_1}$), where $u_1$ is the approach velocity, $g$ is the acceleration of gravity and $d_1$ is the approach flow depth. \textcite{peterka_engineering_1984} established a classical four-regime classification: pre-jumps ($1.7 <$ Fr$_1 < 2.5$), transition jumps ($2.5 <$ Fr$_1 < 4.5$), stabilized jumps ($4.5 <$ Fr$_1 < 9$), and choppy jumps (Fr$_1 > 9$). More recent studies~\cite{chanson_characteristics_1995, valero_airwater_2024} simplify this into three regimes: undular jumps (Fr$_1 \approx 1-2.5$) with minimal air entrainment; transition jumps (Fr$_1 \approx 2.5-4.5$) with partial air entrainment; and strong turbulent jumps (Fr$_1 > 4.5$) with intense air entrainment, fully developed rollers and considerable energy dissipation.

%Temporal Dynamics and Oscillatory Behavior
Beyond their time-averaged characteristics, hydraulic jumps exhibit quasi-periodic oscillations (1--4 Hz) in toe position and free surface elevation, linked to the formation, growth, and advection of large-scale coherent structures within the shear layer, with strong correlations observed among toe oscillations, roller surface fluctuations, and downstream advection of large-scale vortical structures (referred to as vortex ejection in the literature)~\cite{murzyn_free_2007, wang_turbulence_2014, zhang_turbulence_2013}. Numerical prediction of these features remains challenging: while recent Unsteady Reynolds-Averaged Navier-Stokes (URANS) studies have captured toe oscillation frequencies~\cite{bayon-barrachina_numerical_2015}, only a limited number of high-fidelity computational studies have successfully reproduced the complete unsteady behavior~\cite{jesudhas_modelling_2020, mortazavi_direct_2016}.

%Air Entrainment Mechanism
Air entrainment in hydraulic jumps occurs through two primary mechanisms. First, it can take place at the jump toe, where the supercritical jet creates a free surface concavity, entrapping air pockets that are subsequently broken into bubbles~\cite{chanson_experimental_1997, chanson_bubbly_2007}. Second, air is entrained through the roller surface due to free-surface fluctuations and deformations~\cite{wang_turbulence_2014, wang_self-similarity_2016}. The void fraction distribution resulting from bubble transport, breakup and coalescence, typically exhibits a Gaussian profile within the shear layer (Figure~\ref{fig:sketch}), with maximum values decreasing exponentially in the streamwise direction~\cite{chanson_experimental_2000, wang_turbulence_2014}.

%Numerical Modeling Approaches: State of the Art
Numerical simulation of self-aerated flows can be broadly categorized into two groups~\cite{zabaleta_numerical_2024}: (a) bubble-resolved methods where air-water interfaces are discretized with numerous cells, including the dynamics of individual bubbles (typically employing Direct Numerical Simulation (DNS), Large Eddy Simulation (LES), or Detached Eddy Simulation (DES)), and (b) averaged approaches utilizing Eulerian or Lagrangian frameworks to simulate bubble transport (primarily relying on Reynolds-Averaged Navier-Stokes (RANS) and URANS closures). The former requires very fine grids with computational costs that grow rapidly with Reynolds and Weber number ($We=\rho u^2 L / \sigma$, where $\rho$ is the fluid density, $u$ represents a characteristic velocity, $L$ is a characteristic length, and $\sigma$ denotes the surface tension coefficient). In the inertia-dominated turbulent regime, the required number of grid points ($N$) scales as $N \sim We_L^{9/5}$, leading to a total computational cost increase of approximately 240 times for each order of magnitude increase in Weber number~\cite{hatashita_scalings_2025}. These scaling requirements make bubble-resolved simulations practically unfeasible for large-scale engineering applications. The latter offer significantly lower computational cost but relatively reduced spatial resolution~\cite{zabaleta_numerical_2024}.

RANS and URANS models remain the primary tool for laboratory or field hydraulic jumps due to their computational affordability. For mean single-phase features, such as velocity profiles, pressure distributions, and jump geometry, these approaches demonstrate remarkable accuracy, with most mean flow variables achieving errors below 10\%; however, some variables such as roller length and recirculation zone velocities may exhibit differences of 20\% or higher~\cite{bayon_performance_2016}. Turbulence closure selection significantly influences results, and RNG $k$-$\varepsilon$ models have often outperformed standard $k$-$\varepsilon$ and $k$-$\omega$ SST variants~\cite{bayon-barrachina_numerical_2015}. Inherent RANS limitations emerge when predicting air entrainment: The time-averaging inherent to RANS cannot capture the free-surface fluctuations and interfacial instabilities responsible for air entrainment, necessitating additional sub-grid, air-entrainment models with empirical closures and variable performance~\cite{viti_numerical_2018}.% Whereas scale-resolving methods (LES, DES, DNS) of hydraulic jumps address these limitations by explicitly capturing turbulent structures and free-surface dynamics, their computational demands remain prohibitive for routine engineering studies.

Sub-grid approaches that correlate entrainment rates with local turbulence conditions used in hydraulic jump simulations, while computationally efficient, tend to underestimate air concentrations in the roller region and require empirical calibration~\cite{ma_comprehensive_2011}. The fundamental challenge lies in the strong coupling among air entrainment, turbulence generation, and bubble transport~\cite{zabaleta_numerical_2024}. Consequently, models capable of reliably predicting air entrainment across a wide range of Froude numbers with reasonable computational cost remain limited, and systematic validations are scarce.

%Within RANS and URANS frameworks applied to hydraulic jumps, air entrainment has been modeled through two main approaches. Sub-grid models that correlate entrainment rates with local turbulence conditions~\cite{ma_comprehensive_2011}. Whereas computationally efficient, they tend to underestimate air concentrations in the roller region and require empirical calibration~\cite{ma_comprehensive_2011}. Bubble-resolved approaches~\cite{witt_simulating_2015}, by contrast, can achieve entrainment predictions within 10\% of experiments without calibration parameters~\cite{witt_simulating_2015}, but their computational requirements remain prohibitive for most engineering applications. The fundamental challenge lies in the strong coupling among air entrainment, turbulence generation, and bubble transport-physical processes that existing RANS closures struggle to represent comprehensively~\cite{zabaleta_numerical_2024}. As a result, models capable of reliably predicting air entrainment in hydraulic jumps across a wide range of Froude numbers with reasonable computational cost remain limited, and systematic validations are scarce.
%Second, bubble-resolved simulations using Volume of Fluid (VOF) methods have achieved total entrainment predictions within 10\% of experiments without calibration parameters~\cite{witt_simulating_2015}. However, these approaches require have considerable computational costs, as discussed previously.

This study employs the three-phase mixture model recently presented by Zabaleta et al.~\cite{zabaleta_novel_2023, zabaleta_numerical_2024}, which represents the flow as a continuous mixture of three phases: continuous air, continuous water, and dispersed bubbles. This formulation has been successfully validated for stepped spillways and plunging jets~\cite{zabaleta_numerical_2024}, motivating its extension to hydraulic jump configurations as examined here.

The primary contributions of this work include:
\begin{enumerate}
    \item The first comprehensive application of a novel three-phase mixture model to hydraulic jumps across a wide Froude number range (Fr$_1 = 1.98-8.48$), from undular to strong jumps. Direct comparisons with DES~\cite{jesudhas_modelling_2020} and bubble-resolved approaches~\cite{witt_simulating_2015} demonstrate comparable accuracy while requiring substantial reductions in computational resources.
    \item A detailed evaluation of three turbulence closures (RNG $k$-$\varepsilon$, standard $k$-$\varepsilon$, and $k$-$\omega$ SST) and the effect of the turbulent Schmidt number. This study reveals fundamental interactions between turbulence modeling and bubble transport that are critical for accurate air entrainment prediction.
    \item A comprehensive analysis of quasi-periodic jump toe and free-surface oscillations, also validated against different laboratory tests.
    %\item An interpretation of the similarities and differences of flow and air entrainment across the investigated incident Froude number range.
\end{enumerate}

The remainder of this paper is organized as follows: In Section~\ref{sec:NumFramework}, we describe the three-phase mixture framework and implementation details; in Section~\ref{sec:Comp_setup}, the validation methodology and computational setup are presented; in Section~\ref{sec:turbulence_closure}, the turbulence closure and 
turbulent Schmidt number influence on model performance are examined; in Section~\ref{sec:Results}, comprehensive validation results including velocity and air concentration profiles and dynamic behavior are presented; and in Section~\ref{sec:Conclusions}, practical engineering guidelines and future research directions are synthesized.

%% file: II_Theoretical_Model_and_Numerical_Implementation.tex
\section{Theoretical Framework and Numerical Implementation}\label{sec:NumFramework}

\subsection{Three-Phase Mixture Model Overview}
The three-phase mixture formulation employed in this study, developed by Zabaleta et al.~\cite{zabaleta_novel_2023, zabaleta_numerical_2024}, represents self-aerated flows as continuous air flowing over a water–bubble mixture. This approach offers two significant advantages particularly relevant to hydraulic jump simulation: (1) The model accounts for the volume occupied by bubbles in the water-bubble mixture, inherently capturing the increase in water depth due to the presence of bubbles (bulking); and (2) it naturally handles the transition from clear water flow (upstream supercritical region) to highly aerated flow (jump region) to clear water flow (downstream subcritical region where bubbles have escaped) without requiring special treatment or switching between different models.
The flow field is described using three distinct phases characterized by their volume fractions ($\alpha_k$) and densities ($\rho_k$):
\begin{itemize}
    \item Water phase: $\alpha_w$ and $\rho_w$.
    \item Continuous air phase: $\alpha_a$ and $\rho_a$.
    \item Bubble phase: $\alpha_b$ and $\rho_b = \rho_a$ (assuming that variations due to pressure and temperature are negligible).
\end{itemize}
A fourth phase, the water–bubble mixture, is defined as $\alpha_m = \alpha_w + \alpha_b$. This definition allows the interface between the water-bubble mixture and continuous air to be resolved using standard interface-capturing techniques, such as the Volume of Fluid (VOF) method.

The three-phase mixture properties are computed through volume and mass averaging:
\begin{align}
    \rho &= \sum_k \alpha_k \rho_k \quad \text{(density)} \\
    \mathbf{v} &= \frac{1}{\rho} \sum_k \alpha_k \rho_k \mathbf{v}_k \quad \text{(mass-averaged velocity)} \\
    \mathbf{u} &= \sum_k \alpha_k \mathbf{v}_k \quad \text{(volume-averaged velocity)}
\end{align}

\subsection{Governing Equations}
\subsubsection*{Mass Conservation Laws}
The mass conservation equation for the three-phase mixture accounts for relative motion between phases:
\begin{multline}\label{eq:mass-3phase}
\nabla \cdot \mathbf{v} = \nabla \cdot \left[
\alpha_m(1-\alpha_m)\frac{\rho_m - \rho_a}{\rho}\mathbf{u}_{m-a}\right. \\
\left.+ \alpha_b s \frac{\alpha_w}{\alpha_m + \alpha_b s}
\frac{\rho_m}{\rho}\mathbf{v}_{b-w}
\right]
\end{multline}
where $s = \rho_b/\rho_w - 1$ represents the density difference between bubbles and water, $\mathbf{u}_{m-a}$ is the relative velocity between the water-bubble mixture and continuous air, and $\mathbf{v}_{b-w}$ is the slip velocity between bubbles and water.
The mass conservation equation for the water–bubble mixture is:
\begin{equation}\label{eq:mass-bw}
\frac{\partial \alpha_m}{\partial t} + \nabla \cdot (\alpha_m \mathbf{u}) 
+ \nabla \cdot \left[
\alpha_m(1-\alpha_m) \mathbf{u}_{m-a}
\right] = S_b
\end{equation}
The mass conservation equation for the bubble phase is:
\begin{multline}\label{eq:mass-b}
\frac{\partial \alpha_b}{\partial t} + \nabla \cdot (\alpha_b \mathbf{u}) + \nabla \cdot \Big[\alpha_b\Big((1-\alpha_m)\mathbf{u}_{m-a} \\
+ \frac{\alpha_w}{\alpha_m}\mathbf{v}_{b-w}\Big)\Big] = S_b
\end{multline}

\subsubsection*{Momentum Conservation Law}

The momentum conservation equation for the three-phase mixture is:
\begin{multline}\label{eq:mom-3phase}
    \frac{\partial \rho \mathbf{v}}{\partial t} + \nabla \cdot (\rho \mathbf{v}\mathbf{v}) 
    = -\nabla p_{rgh}
    + \nabla \cdot \left[
    \mu_{\text{eff}}(\nabla\mathbf{v}+\nabla\mathbf{v}^T)\right. \\
    \left.+ \mathbf{T}_{b-w} + \mathbf{T}_{m-a} 
    \right]
    - \mathbf{g} \cdot \mathbf{x} \nabla \rho + \mathbf{f}_{\sigma}
\end{multline}
where $p_{rgh} = p - \rho \mathbf{g} \cdot \mathbf{x}$ is the reduced pressure, $\mu_{\text{eff}} = \mu + \mu_t$ is the effective viscosity, where $\mu$ is the dynamic viscosity and $\mu_t$ is the dynamic eddy viscosity, and $\mathbf{f}_\sigma$ represents the surface tension force computed using the Continuum Surface Force model\cite{brackbill_continuum_1992}. The tensors $\mathbf{T}_{b-w}$ and $\mathbf{T}_{m-a}$ account for momentum transfer due to the relative motion between phases, and their full formulation can be found in Zabaleta et al.\cite{zabaleta_novel_2023}.

\subsection*{Closure Models for Relative Velocities}
Instead of solving separate momentum equations for each phase, algebraic models for the relative velocities are used. The relative velocity between bubbles and water combines buoyancy and turbulent dispersion effects, respectively:
\begin{equation}\label{eq:rel_vel}
    \mathbf{v}_{b-w} = \sqrt{1-\alpha_b}\mathbf{v}^\infty_{b-w} + \frac{D_b}{\alpha_b} \nabla \alpha_b
\end{equation}
The first term follows the formulation of \textcite{ishii_thermo-fluid_2011}. The second term represents the transport of bubbles from regions of high concentration to regions of low concentration. The turbulent diffusion coefficient $D_b = \nu_t/Sc_t$ represents bubble diffusion due to turbulent fluctuations, where $\nu_t$ is the kinematic eddy viscosity and $Sc_t$ is the turbulent Schmidt number. The term $S_b$ models air entrainment and degassing processes, as detailed in Section~\ref{sec:entrainment-degassing}.

The terminal slip velocity $\mathbf{v}^\infty_{b-w}$ is calculated using the model from W\"{u}est et al.\cite{wuest_bubble_1992}:
\begin{equation}\label{eq:terminal-velocity}
v^{\infty}_{b-w} = 
\begin{cases}
    -4474 r_b^{1.357} \hat{\mathbf{g}} & \text{if } 0 < r_b \leq 7 \times 10^{-4} \text{ m} \\
    -0.23 \hat{\mathbf{g}} & \text{if } 7 \times 10^{-4} < r_b \leq 5.1 \times 10^{-3} \text{ m} \\
    -4.202 r_b^{0.547} \hat{\mathbf{g}} & \text{if } r_b > 5.1 \times 10^{-3} \text{ m}
\end{cases}
\end{equation}
where $\hat{\mathbf{g}} = \mathbf{g}/|\mathbf{g}|$ is the unit vector and $r_b$ is the bubble radius.
The artificial compression velocity $\mathbf{u}_{m-a}$, used to prevent numerical diffusion at the interface \cite{deshpande_evaluating_2012}, is calculated as:
\begin{equation}
    \mathbf{u}_{m-a} = C_\alpha |\mathbf{u}| \frac{\nabla \alpha_m}{|\nabla \alpha_m|}
\end{equation}
A value of $C_\alpha = 1$ is used throughout this work.

\subsection*{Air Entrainment and Degassing}\label{sec:entrainment-degassing}
The model incorporates a mass transfer mechanism to simulate air entrainment and degassing. Air entrainment occurs when disturbing energy (primarily associated with turbulent kinetic energy $k$) overcomes surface tension and gravity, transforming continuous air into bubbles within a thin mass-transfer region above the water–bubble mixture interface. Conversely, degassing occurs when bubbles rise to the surface and turbulence is not strong enough to entrain air.
The mass transfer between continuous air and dispersed bubbles is represented by the source term $S_b$:
\begin{equation}\label{eq:Sb}
    S_b = \phi S_b^+ + (1-\phi) S_b^-
\end{equation}
where $S_b^+$ represents the air entrainment rate (transformation of continuous air into bubbles) and $S_b^-$ represents the degassing rate (transformation of bubbles into continuous air). These terms are formulated to achieve a bubble concentration of 0.9 in the mass-transfer region within a short time scale, with appropriate limiters to ensure numerical stability~\cite{zabaleta_numerical_2023}.
The switch function $\phi$ determines whether air is being entrained ($\phi = 1$) or degassed ($\phi = 0$) based on an energy balance criterion proposed by \textcite{hirt_modeling_2012}. This criterion compares the disturbing energy with the stabilizing energy (gravitational and surface tension forces).
The complete mathematical formulation and numerical implementation can be found in Zabaleta et al.~\cite{zabaleta_novel_2023, zabaleta_numerical_2024}

%% file: III_Numerical_model_setup_and_validation.tex
\section{Validation Data and Model Implementation}\label{sec:Comp_setup}
All simulations were performed using OpenFOAM v2312, employing a custom solver based on \texttt{interFoam} with the three-phase mixture formulation of Zabaleta et al.~\cite{zabaleta_novel_2023,zabaleta_numerical_2024}

\subsection{Datasets for Experimental Validation}
The model was validated against twelve experimental cases from seven facilities, summarized in Table~\ref{tab:experimentalSetup}, which includes flow conditions and measured variables for each case. The datasets encompass measurements at multiple cross-sections downstream of the jump toe, of normalized streamwise velocity profiles ($u_x/u_1$), void fraction distributions ($C$), and values of turbulent kinetic energy ($k$). %Data were digitized from published figures with careful attention to preserving measurement accuracy.
\begin{table}[htb]
\caption{\label{tab:experimentalSetup}Summary of experimental conditions and validation datasets from literature. Fr$_1$: incident Froude number, $d_1$: inlet flow depth, $u_1$: inlet velocity. Symbols indicate data availability: $\checkmark$: available; $o$: omitted data$^{\dagger}$; empty: no data.}
\begin{ruledtabular}
\begin{tabular}{ccc|ccc|c} % Vertical lines between columns 3-4 and 6-7
Fr$_1$ & $d_1$ & $u_1$  & \multicolumn{3}{c|}{Validation data} & \multirow{2}{*}{Ref.} \\ 
$[-]$ & $[m]$ & $[m/s]$ & $u_x/u_1$ & $C$ & $k$ & \\ \hline
\addlinespace
1.98 & 0.059 & 1.50 &  & $\checkmark$ &  & \cite{murzyn_optical_2005} \\ 
2.00 & 0.071 & 1.67 & $\checkmark$ & $o$ & $\checkmark$ & \cite{liu_evaluation_2004, liu_turbulence_2004} \\ 
2.10 & 0.097 & 2.10 & $\checkmark$ & $\checkmark$ &  & \cite{wuthrich_hydraulic_2022} \\ 
2.40 & 0.084 & 2.21 & $\checkmark$ & $\checkmark$ & & \cite{wuthrich_hydraulic_2022} \\ 
3.32 & 0.041 & 2.10 & $\checkmark$ & $o$ & $\checkmark$ & \cite{liu_evaluation_2004, liu_turbulence_2004} \\ 
4.80 & 0.021 & 2.19 &  & $\checkmark$ & & \cite{murzyn_optical_2005} \\ 
5.10 & 0.018 & 2.12 & $o$ & $\checkmark$ & & \cite{murzyn_free_2007} \\ 
5.80 & 0.024 & 2.28 &  & $\checkmark$ & & \cite{kucukali_turbulence_2007} \\ 
6.33 & 0.014 & 2.34 & $\checkmark$ & $\checkmark$ & & \cite{chanson_experimental_2000} \\ 
6.90 & 0.024 & 3.35 & $o$ & $\checkmark$ & $\checkmark$ & \cite{kucukali_turbulence_2007} \\ 
7.50 & 0.020 & 3.33 & $\checkmark$ & $\checkmark$ & $\checkmark$ & \cite{wang_turbulence_2014} \\ 
8.48 & 0.014 & 3.14 & $\checkmark$ & $\checkmark$ & & \cite{chanson_experimental_2000} \\ 
\end{tabular}
\end{ruledtabular}
\begin{tablenotes}
\item[]$\dagger$ Velocity data at Fr$_1 = 5.1$ and $6.9$ were discarded for validation due to profiles with insufficient vertical extent or anomalous shape inconsistent with the expected wall-jet structure of hydraulic jumps. Concentration data at Fr$_1 = 2.0$ and $3.32$ were excluded because measured concentrations remain below 8\%, falling within the numerical uncertainty of the simulations.
\end{tablenotes}
\end{table}
%\subsubsection*{Measurement Uncertainties and Validation Considerations}\label{subsec:ExpMeasurementConsiderations}
The validation using experimental measurements in highly aerated flows requires careful consideration of inherent measurement uncertainties. Velocity measurements using dual-tip phase-detection probes exhibit uncertainties ranging from $\pm5\%$ in moderately aerated regions to $\pm10\%$ in extreme concentration zones. Void fraction measurements show accuracies of $\pm2\%$ for intermediate concentrations, but uncertainties increase substantially for $C < 0.05$~\cite{kramer_best_2020}. In this context, our validation strategy employs Mean Absolute Error (MAE) as the primary quantitative metric, acknowledging that inherent measurement uncertainties limit strict point-wise comparisons.

For post-processing and validation purposes, the total air concentration is computed as $C = \alpha_b + \alpha_a$, representing the sum of bubble and continuous air volumetric fractions. This definition accounts for the unsteady nature of hydraulic jumps, where instantaneous flow fields contain regions of continuous air above the fluctuating free surface. The characteristic flow depth $y_{90}$ is defined as the elevation where $C = 0.90$, following standard conventions in air-water flow research~\cite{zabaleta_novel_2023}. The time-averaged jump toe position ($x_{toe}$) throughout this study is defined as the streamwise location where the time-averaged surface elevation first exceeds $1.05 d_1$, marking the onset of flow depth increase from the incoming supercritical jet.

\subsection{Numerical Discretization Schemes}
%\subsubsection*{Spatial and Temporal Discretization}
The numerical discretization schemes employed in the simulations are summarized in Table~\ref{tab:numerical_schemes}.
\begin{table}[htb]
\caption{\label{tab:numerical_schemes}Spatial and temporal discretization schemes employed in this study.}
\begin{ruledtabular}
\begin{tabular}{ll}
\textbf{Term/Variable} & \textbf{Scheme} \\ \hline
Time derivative ($\partial/\partial t$) & Implicit Euler \\
\addlinespace[0.5ex]
\multicolumn{2}{l}{\textit{Convective terms:}} \\
\quad Three-phase mixture ($\nabla \cdot (\rho\mathbf{v}\mathbf{v})$) & Linear upwind \\
\quad Turbulence ($\nabla \cdot (\rho\mathbf{v}k)$, $\nabla \cdot (\rho\mathbf{v}\varepsilon)$) & Linear upwind  \\
\quad Turbulence ($\nabla \cdot (\rho\mathbf{v}\omega)$) & Linear upwind \\
\quad Bubble phase ($\nabla \cdot (\alpha_b\mathbf{v})$) & Upwind \\
\quad Water-bubble mixture ($\nabla \cdot (\alpha_m\mathbf{v})$) & van Leer with $C_\alpha = 1$ \\
\addlinespace[0.5ex]
\multicolumn{2}{l}{\textit{Diffusive terms:}} \\
\quad Gradient ($\nabla \phi$) & Cell-limited Gauss linear \\
\quad Laplacian ($\nabla^2 \phi$) & Gauss linear corrected \\
\end{tabular}
\end{ruledtabular}
\end{table}
The pressure-velocity coupling was solved using the PIMPLE algorithm with $1000$ outer correctors and $2$ pressure correctors per time step. Under-relaxation factors of $0.2$ were applied to pressure and velocity fields during intermediate iterations. The simulation time step was dynamically adjusted to maintain the Courant number below $0.5$.

\subsection{Computational Domain and Mesh Configuration}
\begin{figure*}[htb]
    \centering
    \includegraphics[width=\textwidth]{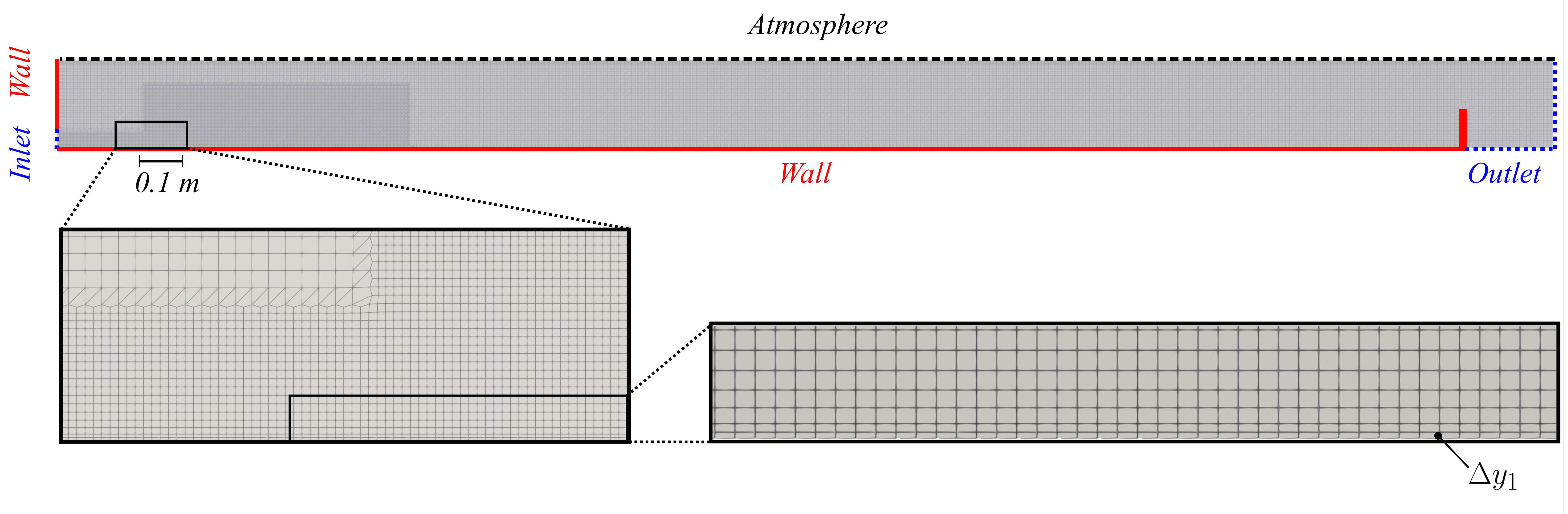}
    \caption{Computational domain, boundary conditions, and mesh configuration. The upper panel shows the full domain with boundary conditions. The lower-left panel details the locally refined mesh at the jump toe region. The lower-right panel shows the mesh close to the wall, with $\Delta y_1$ indicating the first cell height.}
    \label{fig:meshBC}
\end{figure*}

The computational domain replicates experimental hydraulic jump configurations in horizontal rectangular channels, with dimensions matching the experimental conditions as illustrated in Figure~\ref{fig:meshBC}. Each case was modeled as two-dimensional, with an overflow gate at the downstream boundary to fix the tailwater level. The mesh consisted of quadrilateral cells with local refinement in the jump toe and roller region, as illustrated in Figure~\ref{fig:meshBC}. Cell sizes in the refined regions ranged from 10 to 5 mm. Near-wall regions were further refined by subdividing the first 5 mm cell into 2–3 subcells with geometric expansion ratios between 1.1 and 1.5, to attain values of  $y^+$ between 30 and 300. This procedure yielded first-cell heights between 1.2 and 2.6 mm across the simulated cases (Table \ref{tab:mesh_details}). This wall refinement extended the entire horizontal axis of the domain.
\begin{table}[htb]
\caption{\label{tab:mesh_details}Details of the computational mesh and model parameters for all simulated cases.}
\begin{ruledtabular}
\begin{tabular}{ccccccc}
$Fr$ & Grid  & \multicolumn{3}{c}{y+ Statistics} & $d_b$ & $\Delta y_1$ \\
 & Size & Min & Max & Avg & [mm] & [mm] \\ \hline
1.98 & 15,681 & 65 & 187 & 123  & 3.0 & 2.6\\%murzyn05
2.00 & 24,429 & 90 & 203 & 156 & 3.0 & 2.5\\%Liu04
2.10 & 19,511 & 121 & 278 & 211 & 3.0 & 2.1\\
2.40 & 17,720 & 109 & 275 & 185 & 2.8 & 2.0\\
3.32 & 26,545 & 84 & 263 & 149 & 2.5 & 2.0\\
4.80 & 13,271 & 52 & 179 & 95 & 4.0 & 1.6\\
5.10 & 11,393 & 23 & 136 & 61 & 3.0 & 1.8\\
5.80 & 20,067 & 94 & 318 & 184 & 2.0 & 1.5\\
6.33 & 11,899 & 27 & 125 & 51 & 1.5 & 1.5\\
6.90 & 20,979 & 32 & 196 & 115 & 1.0 & 1.2\\
7.50 & 18,449 & 52 & 197 & 94 & 1.0 & 1.2\\
8.48 & 17,026 & 37 & 204 & 110 & 1.0 & 1.2
\end{tabular}
\end{ruledtabular}
\end{table}

\subsubsection*{Boundary and Initial Conditions}
The computational domain employed four boundary types, configured to replicate experimental conditions. At the inlet, we defined uniform velocity profiles ($u = u_1$, $v = 0$) with pure water conditions ($\alpha_w = 1.0$, $\alpha_b = \alpha_a = 0$), to ensure the experimental value of $d_1$ at the toe. This inlet configuration corresponds to partially developed inflow conditions consistent with the experimental validation datasets~\cite{chanson_experimental_2000, murzyn_optical_2005, kucukali_turbulence_2007, wang_turbulence_2014}. Turbulent quantities ($k$, $\varepsilon$ and $\omega$) were prescribed assuming a turbulence intensity of 2\%. 
A sensitivity analysis confirmed that results are insensitive to inlet turbulence intensity within the range Tu = 2–5\% for all cases, with the exception of Fr$_1$ = 2, for which a minor effect was observed in the lower part of the turbulence intensity profile.
At the outlet, we implemented a zero-gradient condition for all transported quantities, whereas the top boundary remained open to atmospheric pressure, allowing bidirectional air exchange. At the bottom walls, we defined no-slip conditions with wall functions for turbulence quantities.

The domain was initialized with two distinct water regions to accelerate relaxation toward quasi-steady conditions: A supercritical flow region with depth $d_1$ extending approximately one meter downstream from the inlet, transitioning to a subcritical depth $d_2$ toward the outlet. The bubble phase fraction was initialized at zero throughout the domain, with air entrainment developing naturally during the simulation.

For each Froude number, a base simulation was first run until the transient was over, and then flow fields from these base runs were used as initial conditions for all subsequent simulations, so that flow statistics were subsequently collected over 30-second intervals with sampling at $0.1$-second increments. Statistical stationarity was confirmed by monitoring volume-integrated bubble fraction $\alpha_b$ and streamwise velocity component $u_x$ (Figure \ref{fig:2007_Murzyn_Fr5.1_stationarity}, Appendix \ref{app:stationarity}). This temporal resolution proved sufficient to capture the characteristic fluctuations of jump toe oscillations and dynamics, as analyzed in Section V.

\subsubsection*{Mesh-Independence and Grid Convergence Analysis}
A mesh-independence analysis was conducted and the Grid Convergence Index (GCI) methodology of Celik et al.~\cite{celik_procedure_2008} was applied for a representative case at Fr$_1 = 8.48$. Three successive refinement levels were evaluated, fine (3.5 mm), medium (5 mm), and coarse (7 mm), with refinement ratio $r \approx \sqrt{2}$, keeping the near-wall mesh refinement unchanged. Table~\ref{tab:gci_results} presents the GCI analysis for roller length ($L_r/d_1$),  spatially-averaged peak streamwise velocity ($\overline{u}_{\max}/u_1$), and spatially averaged peak air concentration in the shear layer ($\overline{C}_{\max}^{\rm shear}$), where overbars denote averages over three downstream cross-sections. The apparent order of accuracy $p$, calculated from the solution differences across grid levels~\cite{celik_procedure_2008}, ranges from 2.4 to 3.0. Grid convergence uncertainties (GCI$_{\text{fine}}$) remained below 1.7\% for all reported variables, confirming negligible numerical uncertainty. The computational cost of the fine mesh was approximately three times the cost of the medium mesh. These results justify the selection of the medium mesh resolution for all subsequent simulations. Representative convergence plots showing free surface and roller evolution for the three meshes, along with error plots are provided in Appendix~\ref{app:Mesh}.
\begin{table}[htb]
\centering
\caption{\label{tab:gci_results}Grid Convergence Index analysis for Fr$_1 = 8.48$ following Celik et al.~\cite{celik_procedure_2008}. The medium mesh ($\Delta x = \Delta y = 5$ mm) was selected for all subsequent simulations.}
\begin{ruledtabular}
\begin{tabular}{lccccc}
\textbf{Variable} & Fine & Medium & Coarse & $p$ & GCI$_{\text{fine}}$ \\
& (3.5 mm) & (5 mm) & (7 mm) & [-] & [\%] \\
\midrule
$L_r/d_1$                        & 44.28 & 43.79 & 45.07 & 2.77 & 0.86 \\
$\overline{u}_{\max}/u_1$        & 0.882 & 0.867 & 0.831 & 2.41 & 1.68 \\
$\overline{C}_{\max}^{\rm shear}$& 0.591 & 0.586 & 0.587 & 3.03 & 0.58
\end{tabular}
\end{ruledtabular}
\end{table}

Table~\ref{tab:mesh_details} summarizes the final mesh configurations and computational parameters for each simulation. Total number of cells ranged from approximately 11,400 to 26,500, with variation primarily driven by differences in jump downstream conjugate depth ($d_2$). Computational costs for the 30-second post-transient averaging period ranged from approximately 0.75 to 1.8 CPU-h using 7 processors, with no systematic dependence on Froude number.

\subsection{Representative Bubble Diameter Selection}
The representative bubble diameter ($d_b$) influences bubble dynamics through the slip velocity (Equation~\eqref{eq:terminal-velocity}). Experimental studies consistently report bubble size distributions characterized by log-normal probability density functions~\cite{chanson_air_1997}, with characteristic diameters ranging from 0.5 to 10 mm depending on Froude number and location within the jump. Liu et al.~\cite{liu_evaluation_2004} documented bubble diameter distributions with peak frequencies at $2-3$ mm for Fr$_1 = 2-3.32$. Murzyn et al.~\cite{murzyn_optical_2005} reported Sauter mean diameters of $2-10$ mm for Fr$_1 = 2-6.3$. Kucukali and Chanson~\cite{kucukali_turbulence_2007} found bubble chord time distributions indicating characteristic sizes of 0.5--2 mm for Fr$_1 = 6.9$.
For each simulation, representative diameters were selected (Table~\ref{tab:mesh_details}) based on experimental data at matching Froude numbers when available, or through interpolation following the reported density function.

%% file: IV_Validation_and_parameter_selection.tex
\section{Selection of the Turbulence Closure and Turbulent Schmidt Number}\label{sec:turbulence_closure}
The accurate prediction of air entrainment and bubble distribution in self aerated flows within the three-phase mixture approach depends strongly on the choice of turbulence closure model~\cite{zabaleta_numerical_2024}, which determines the spatial distribution and magnitude of the eddy viscosity ($\nu_t$), which in combination with the turbulent Schmidt number $Sc_t$, defines the turbulent bubble diffusivity $D_b = \nu_t/Sc_t$. This section evaluates the performance of three widely used RANS turbulence closures: standard $k$-$\varepsilon$, RNG $k$-$\varepsilon$, and $k$-$\omega$ SST.

\subsection{Turbulence Closure Models: Performance Assessment}
Each of these two-equation RANS closures has been extensively used for the simulation of hydraulic jump flows. The standard $k$-$\varepsilon$~\cite{launder_application_1974} has been widely applied to hydraulic jumps~\cite{chippada_numerical_1994, gonzalez_two-phase-flow_2005, abbaspour_numerical_2009, ebrahimi_numerical_2013, bayon-barrachina_numerical_2015, harada_modelling_2018}; the RNG $k$-$\varepsilon$~\cite{yakhot_development_1992} incorporates additional terms designed for rapidly strained flows, with demonstrated success in hydraulic jump applications~\cite{carvalho_numerical_2008, abbaspour_numerical_2009, bayon-barrachina_numerical_2015, witt_simulating_2015}. The $k$-$\omega$ SST~\cite{menter_24th_1993} uses $k$-$\omega$ formulation in the near-wall region and $k$-$\varepsilon$ in the outer flow, and has also been used to simulate hydraulic jumps~\cite{ma_modeling_2011, bayon-barrachina_numerical_2015}.
%All models were implemented with variable-density formulations following Zabaleta et al.~\cite{zabaleta_numerical_2024} to prevent unphysical turbulent energy transfer across the large air-water density ratio. However, as discussed in Section~\ref{sec:schmidt_number}, the intense localized air entrainment at the jump toe creates conditions where this protection mechanism becomes affected, necessitating an adjusted turbulent Schmidt number.

\subsubsection*{Turbulent Kinetic Energy}
\begin{figure}[htb]
\centering
\includegraphics[width=\columnwidth]{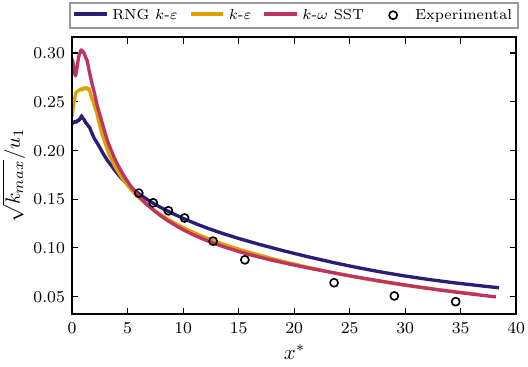}
\caption{Streamwise evolution of the square root of the maximum turbulent kinetic energy normalized by upstream velocity at Fr$_1 = 2$. Experimental data from Liu et al.~\cite{liu_turbulence_2004} (symbols).}
\label{fig:tke05_U1}
\end{figure}

In Figure~\ref{fig:tke05_U1}, we present the streamwise evolution of maximum turbulent kinetic energy normalized by the upstream velocity for Fr$_1 = 2$, comparing model predictions against experimental measurements from Liu et al.~\cite{liu_turbulence_2004}, at different non-dimensional streamwise locations ($x^* = (x-x_\mathrm{toe})/d_1$).

All three closures exhibit a near-toe peak ($x^* < 5$) where experimental data are unavailable. In this unobserved region, the closures diverge considerably: $k$-$\omega$ SST reaches the highest peak ($\sqrt{k_\mathrm{max}}/u_1 
\approx 0.30$), followed by standard $k$-$\varepsilon$ ($\approx 0.26$), with RNG $k$-$\varepsilon$ producing the lowest peak ($\approx 0.23$). Beyond $x^* \approx 5$, all three models are close and reproduce the experimental decay trend through approximately $x^* \approx 12$.  At larger distances ($x^* > 12$), the trend is marked by some overestimation, with the RNG $k$-$\varepsilon$ exhibiting the highest values.

\subsubsection*{Turbulence Intensity}

\begin{figure}[htb]
\centering
\captionsetup[subfigure]{aboveskip=0pt, belowskip=0pt}

% (a)
\begin{subfigure}[b]{\columnwidth}
    \centering
    \includegraphics[width=1\columnwidth]{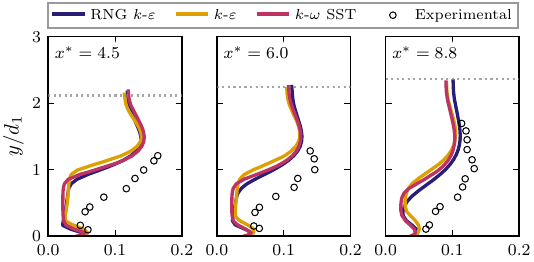}
    \caption{}\label{fig:2004_Liu_Fr2_turbulenceClosure_Tu}
\end{subfigure}

% (b)
\begin{subfigure}[b]{\columnwidth}
    \centering
    \includegraphics[width=1\columnwidth]{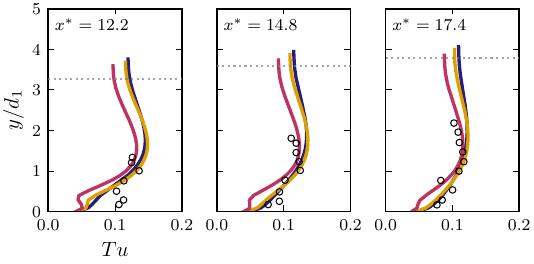}
    \caption{}\label{fig:2004_Liu_Fr3.32_turbulenceClosure_Tu}
\end{subfigure}
\caption{Comparison between simulated (lines) and experimental~\cite{liu_turbulence_2004} (open symbols) turbulence intensity profiles ($Tu= \sqrt{(2/3)k}/u_1$) at three downstream positions.
(a) Fr$_1 = 2$,  
(b) Fr$_1 = 3.32$. Gray dotted line represents experimental $y_{90}$}
\label{fig:all_TC_Tu_profiles}
\end{figure}

In Figures~\ref{fig:2004_Liu_Fr2_turbulenceClosure_Tu} and~\ref{fig:2004_Liu_Fr3.32_turbulenceClosure_Tu}, we compare vertical profiles of turbulence intensity $Tu = \sqrt{(2/3)k}/u_1$ at selected cross-sections against experimental  data from Liu et al.~\cite{liu_turbulence_2004}. Experimental values were reconstructed from reported streamwise ($\overline{u_x'^2}$) and vertical ($\overline{u_y'^2}$) fluctuations. Since the spanwise component ($\overline{u_z'^2}$) was not reported, it was estimated based on the measurements of for open-channel flows by Nezu and Nakagawa\cite{Nezu1993, bombardelli_hierarchical_2009}, $\overline{u_z'^2} = 0.5\overline{u_x'^2}$, yielding $k = 0.5(\overline{u_x'^2} + \overline{u_y'^2}+0.5\overline{u_x'^2})$.

The three closures produce nearly identical turbulence intensity distributions throughout the flow depth and across all downstream positions for both Fr$_1$, indicating limited sensitivity of turbulence intensity to the specific turbulence closure formulation. For Fr$_1 = 2$ (Figure~\ref{fig:2004_Liu_Fr2_turbulenceClosure_Tu}), experimental data are available only within the lower part of the water column ($y/d_1 < 1.5$). In this region, all models underpredict the measured turbulence intensity for all cross-sections. Agreement improves at the furthest measured location ($x^* = 8.8$), where model predictions approach the experimental values.

For Fr$_1 = 3.32$ (Figure~\ref{fig:2004_Liu_Fr3.32_turbulenceClosure_Tu}), experimental data span barely the lower-half of the flow depth ($y/d_1 < 2$). Whereas underprediction close to the wall is present for the first profile ($x^* = 12.2$), all the model results are close to the experimental values at mid-depth ($y/d_1 \approx 1$--$2$).

\begin{figure}[htb]
\centering
\includegraphics[width=1\columnwidth]{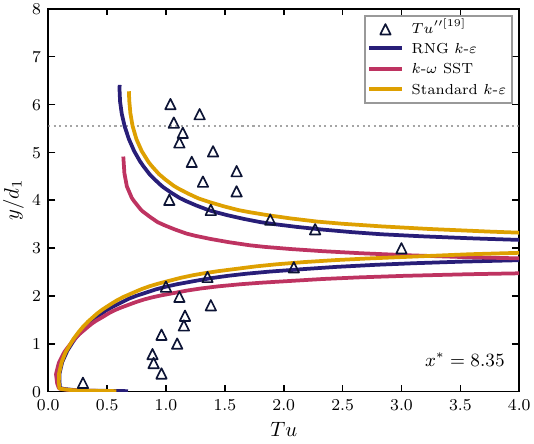}
\caption{Comparison of turbulence intensity profiles at $x^* = 8.35$ for Fr$_1 = 7.5$. Symbols: high-frequency turbulence intensity $Tu''$ from Wang et al.~\cite{wang_experimental_2014}. Lines: numerical results calculated as $Tu = \sqrt{(2/3)k}/|\mathbf{v}|$ from time-averaged fields. Gray dotted line indicates experimental $y_{90}$.}
\label{fig:Wang_Fr7.5_Tu_decomposition}
\end{figure}

Turbulence intensity predictions are compared against frequency-decomposed experimental data from Wang et al.~\cite{wang_experimental_2014} for Fr$_1 = 7.5$ in Figure~\ref{fig:Wang_Fr7.5_Tu_decomposition}. The experimental measurements employed a triple decomposition technique~\cite{felder_triple_2014} to separate the raw turbulence intensity signal ($Tu$) into low-frequency ($Tu'$) and high-frequency ($Tu''$) components. The low-frequency component captures pseudo-periodic fluctuations associated with global jump instabilities including toe oscillations, free-surface distortion, and large-scale vortex advection, while the high-frequency component represents small-scale turbulence~\cite{wang_experimental_2014}.
In this case, turbulence intensity is normalized by the local mean velocity magnitude $|\mathbf{v}|$, yielding $Tu = \sqrt{(2/3)k}/|\mathbf{v}|$, consistent with the interfacial velocity used in  the phase-detection probe measurements~\cite{wang_experimental_2014}. 

In URANS simulations, the turbulent kinetic energy $k$ represents modeled fluctuations based on the Reynolds-averaging. Therefore, calculated $Tu$ corresponds to the high-frequency component $Tu''$. Near the stagnation streamline, located between $2 < y/d_1 < 3.5$ (depending on the turbulence closure), $|\mathbf{v}| = 0$, causing $Tu$ to asymptotically approach infinity.
%, while larger-scale unsteady motions such as jump toe oscillations can be resolved by the time-dependent equations~\cite{viti_numerical_2018}

Whereas all profiles share a similar shape, those pertaining to the $k$-$\varepsilon$ variants are shifted upward in the vertical direction relative to $k$-$\omega$ SST. In the lower shear flow region ($y/d_1 \lesssim 2$), all three closures underpredict $Tu''$. At mid-depth, agreement with experimental data improves, with $k$-$\omega$ SST showing the closest agreement with data.

\subsubsection*{Air Concentration}
\begin{figure}[htb]
\centering
\captionsetup[subfigure]{aboveskip=0pt, belowskip=0pt}

\begin{subfigure}[b]{\columnwidth}
    \centering
    \includegraphics[width=1\columnwidth]{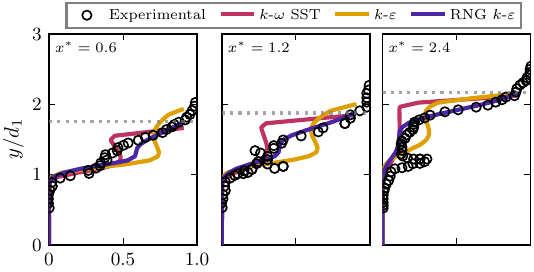}
    \caption{}\label{fig:2022_Wutrich_Fr2.4_TC_C}
\end{subfigure}

\begin{subfigure}[b]{\columnwidth}
    \centering
    \includegraphics[width=1\columnwidth]{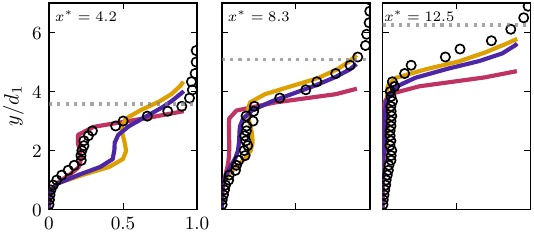}
    \caption{}\label{fig:2007_Murzyn_Fr5.1_TC_C}
\end{subfigure}

\begin{subfigure}[b]{\columnwidth}
    \centering
    \includegraphics[width=1\columnwidth]{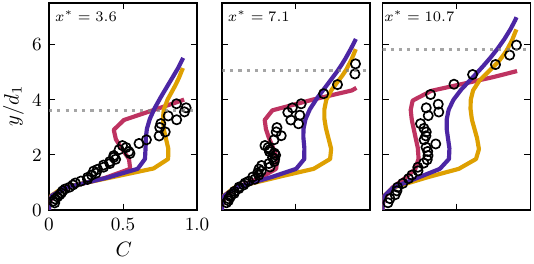}
    \caption{}\label{fig:2000_CB_Fr8.48_TC_C}
\end{subfigure}
\caption{Comparison of vertical concentration profiles for RNG~$k$-$\varepsilon$ ($Sc_t = 3$), standard~$k$-$\varepsilon$ ($Sc_t = 3$), and $k$-$\omega$ SST ($Sc_t = 1$) at three downstream stations:
(a)~Fr$_1 = 2.4$,
(b)~Fr$_1 = 5.1$,
(c)~Fr$_1 = 8.48$.}
\label{fig:TC_concentration_profiles}
\end{figure}

Profiles of air concentration provide a comprehensive test of turbulence closure performance, as they integrate the effects of turbulence production, eddy viscosity, and bubble diffusion~\cite{zabaleta_numerical_2024}. Comparisons at three Froude numbers are presented in Figure~\ref{fig:TC_concentration_profiles}, with MAE values for the remaining cases provided in Table~\ref{tab:turbulence_closure_performance}.
At low Froude numbers (Figure~\ref{fig:2022_Wutrich_Fr2.4_TC_C}), differences among closures are modest and all reproduce the characteristic S-shaped concentration profile, with Mean Absolute Error (MAE) values in the range 0.05--0.11 for all closures (Table~\ref{tab:turbulence_closure_performance}).

At intermediate Froude numbers (Figure~\ref{fig:2007_Murzyn_Fr5.1_TC_C}), both $k$-$\varepsilon$ variants overpredict the first station concentrations but capture the concentration peak and downstream decay for the other two stations. The $k$-$\omega$ SST model reproduces the first measurement station well but underpredicts concentrations further downstream, yielding comparable MAE values of 0.08–-0.11 (Table~\ref{tab:turbulence_closure_performance}).

At high Froude numbers (Figure~\ref{fig:2000_CB_Fr8.48_TC_C}), $k$-$\omega$ SST achieves the closest overall agreement (MAE $= 0.10$), while both $k$-$\varepsilon$ variants overpredict concentrations throughout the flow depth, with RNG (MAE $= 0.11$) performing better than standard $k$-$\varepsilon$ (MAE $= 0.17$).

In Table~\ref{tab:turbulence_closure_performance}, we summarize MAE across ten cases spanning $1.98 \le \text{Fr}_1 \le 8.48$. $k$-$\omega$ SST achieves the lowest MAE in seven of ten cases, while RNG~$k$-$\varepsilon$ performs best at low-to-intermediate Froude numbers, achieving the lowest MAE in three cases. The standard $k$-$\varepsilon$ model is consistently the closure with lowest agreement with data. Overall, $k$-$\omega$ SST and RNG~$k$-$\varepsilon$ produce comparable accuracy across the full range, with MAE differences of $0.01$--$0.03$ between the two. 

The observed performance is consistent with the known properties of each closure~\cite{wilcox_turbulence_1998, pope_turbulent_2000, bridgeman_computational_2009}, and illustrates the flow-dependence of RANS closure selection across hydraulic applications~\cite{sibil_comparison_2021}. The standard $k$-$\varepsilon$ model's deterioration with Fr$_1$ reflects its known limitations in flows with strong curvature and adverse pressure gradients~\cite{bridgeman_computational_2009}. The superior performance of SST at high Fr$_1$ aligns with the findings of Zabaleta et al.\cite{zabaleta_numerical_2024} for plunging jets, which share some analogous air entrainment mechanisms with hydraulic jumps\cite{chanson_air_1995,chanson_air_1995-1}.

\begin{table}[htb]
\caption{\label{tab:turbulence_closure_performance}Mean Absolute Error (MAE) for concentration predictions averaged across all measurement positions within each case. Bold indicates lowest values. Schmidt number values: 
$Sc_t = 1$ for $k$-$\omega$ SST, $Sc_t = 3$ for both $k$-$\varepsilon$ variants.}
\begin{ruledtabular}
\begin{tabular}{lccc}
 & \multicolumn{3}{c}{MAE} \\
Fr$_1$
 & $k$-$\omega$ SST & RNG $k$-$\varepsilon$ & Standard $k$-$\varepsilon$ \\
\hline
1.98 & \textbf{0.10} & 0.11 & 0.10 \\
2.10 & 0.07 & \textbf{0.05} & 0.06 \\
2.40 & 0.09 & \textbf{0.05} & 0.11 \\
4.80 & \textbf{0.09} & 0.13 & 0.11 \\
5.10 & \textbf{0.07} & 0.08 & 0.08 \\
5.80 & \textbf{0.09} & 0.11 & 0.14 \\
6.33 & 0.15 & \textbf{0.13} & 0.16 \\
6.90 & \textbf{0.09} & 0.10 & 0.16 \\
7.50 & \textbf{0.08} & 0.12 & 0.16 \\
8.48 & \textbf{0.10} & 0.11 & 0.17
\end{tabular}
\end{ruledtabular}
\end{table}

\subsection{Turbulent Schmidt Number Selection}\label{sec:schmidt_number}
For bubbly flows, values of $Sc_t$ ranging $0.7$--$1.2$ are commonly adopted~\cite{gualtieri_values_2017}. However, in hydraulic jumps, the intense localized air entrainment at the jump toe reduces mixture density from $\rho \approx 1000$~kg/m$^3$ to $\rho \approx 250$--$500$~kg/m$^3$, weakening the density-based separation that confines turbulence to each phase in variable-density RANS formulations~\cite{zabaleta_numerical_2024}, increasing $k$ and consequently $\nu_t$ beyond physically realistic levels. This leads to excessive bubble diffusivity ($D_b = \nu_t/Sc_t$) and overpredicted downstream air concentrations.

\begin{figure}[htb] 
\centering \includegraphics[width=\columnwidth]{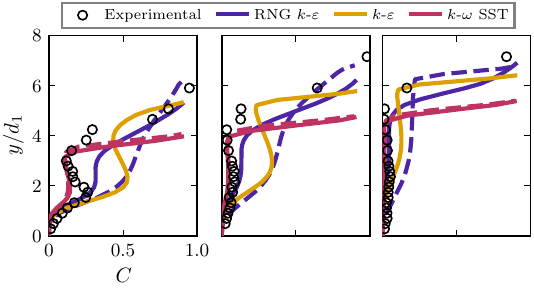}
\caption{Vertical concentration profiles at Fr$_1 = 5.8$ for the three closures ($Sc_t = 1$ dashed, $Sc_t = 3$ solid), at three downstream stations.} \label{fig:Schmidt_concentration_profiles}
\end{figure}

This effect varied for the two closure families. We suggest that, for $k$-$\varepsilon$ variants, an overprediction of $k$ at the toe propagates through the jump due to the strong coupling between $k$ and its production rate. The artificially high $k$ at the toe increases the modeled turbulent viscosity ($\nu_t \propto k^2/\varepsilon$), which in turn over-predicts the generation of turbulence within the highly sheared roller region~\cite{durbin_k-3_1996}. With $Sc_t = 1$, the resulting $D_b$ is excessive, producing severe concentration overprediction at intermediate and high Froude numbers (Figure \ref{fig:Schmidt_concentration_profiles}). Adopting $Sc_t = 3$, consistent with the observed $2-4\times$ density reduction at the toe, counteracts this inflation and restores a physically realistic diffusion field, as illustrated in Figure~\ref{fig:Schmidt_concentration_profiles}. It is important to note that $Sc_t = 3$ should not be interpreted as a totally  physical value; it compensates for the overprediction of $\nu_t$ by the $k$-$\varepsilon$ turbulence models in this flow configuration.

In contrast, the $k$-$\omega$ SST model mitigates this spurious behavior through its modified eddy viscosity formulation and built-in production limiters. By inherently bounding the turbulent shear stress in high-strain regions like the jump toe, the model prevents the self-reinforcing overproduction of $k$. Because the SST model avoids artificially increasing the turbulent viscosity, it does not spuriously exaggerate bubble mixing. As a consequence, it is practically insensitive to the turbulent Schmidt number: concentration profiles for $Sc_t = 1$ and $Sc_t = 3$ are nearly coincident across the full Froude number range studied (Table~\ref{tab:turbulence_closure_performance}), confirming that $Sc_t = 1$ is an appropriate and more physically grounded choice.

In Figure~\ref{fig:Schmidt_concentration_profiles}, we illustrate the bubble concentration at Fr$_1 = 5.8$. For RNG~$k$-$\varepsilon$ with $Sc_t = 1$, the model overpredicts $C$ throughout the flow depth; adopting $Sc_t = 3$ corrects the profiles significantly. For $k$-$\omega$ SST, profiles for $Sc_t = 1$ and $Sc_t = 3$ are nearly identical, confirming the weak sensitivity discussed above.

Considering the comparable accuracy of RNG~$k$-$\varepsilon$ and $k$-$\omega$ SST, the fact that $k$-$\omega$ SST performs well with the standard $Sc_t = 1$, and its demonstrated robustness in related aerated flow configurations~\cite{zabaleta_numerical_2024}, $k$-$\omega$ SST with $Sc_t = 1$ is selected for all subsequent simulations in Section~\ref{sec:Results}.

%% file: V_Results.tex
\section{Results}\label{sec:Results}
This section presents comprehensive validation of the three-phase mixture model across the twelve experimental datasets. $k$-$\omega$ SST was used with $Sc_t = 1$.

\begin{figure*}[htb]
\includegraphics[width=1\textwidth]{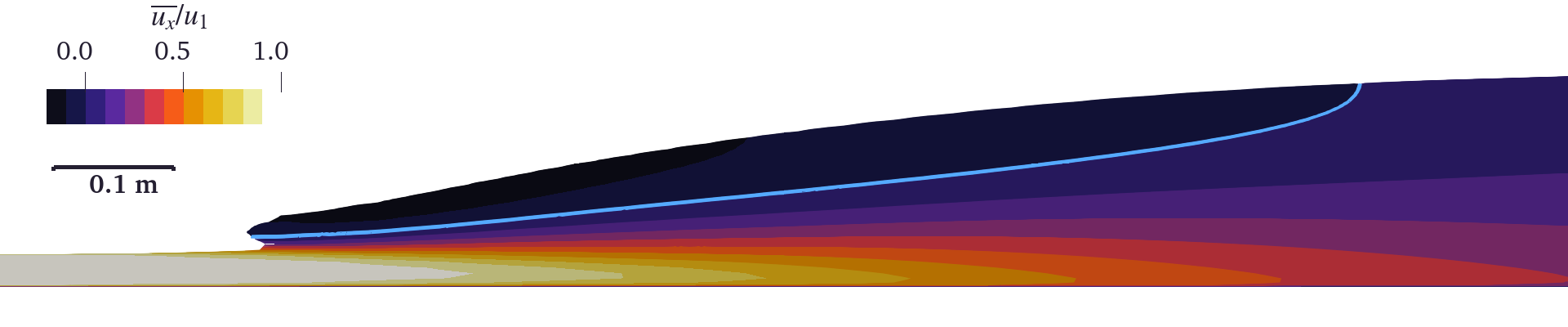}
\caption{\label{fig:velocityField_Fr5.8}Normalized mean streamwise velocity field for Fr$_1 = 5.8$. Cyan line indicates the mean stagnation streamline.}
\end{figure*}

\subsection{Velocity Profiles}\label{subsec:velocity}
The normalized velocity field for Fr$_1 = 5.8$ is presented in Figure~\ref{fig:velocityField_Fr5.8}. The cyan line indicates the mean stagnation streamline. The different flow regions are visible, with an incoming high velocity jet, the turbulent shear layer separating the jet and the roller with intense velocity gradients, the roller region with negative velocities, and a downstream low velocity in the uniform condition.

Table~\ref{tab:MAEVelocity} quantifies predictive performance through MAE computed on normalized velocity profiles ($u_x/u_1$) at all measured downstream positions for each case.

\begin{table}[htb]
\centering
\caption{Mean Absolute Error (MAE) for normalized streamwise velocity profiles ($u_x/u_1$), averaged across all measured cross sections for each case.}
\label{tab:MAEVelocity}
\begin{ruledtabular}
\begin{tabular}{c*{7}{c}}
Fr$_1$ & 2.00 & 2.10 & 2.40 & 3.32 & 6.33 & 7.50 & 8.48 \\
\hline
MAE & 0.04 & 0.31 & 0.27 & 0.07 & 0.23 & 0.33 & 0.21 \\
\end{tabular}
\end{ruledtabular}
\end{table}

\begin{figure}[htbp]
\centering
\captionsetup[subfigure]{aboveskip=0pt, belowskip=0pt}

\begin{subfigure}[b]{\columnwidth}
    \centering
    \includegraphics[width=1\columnwidth]{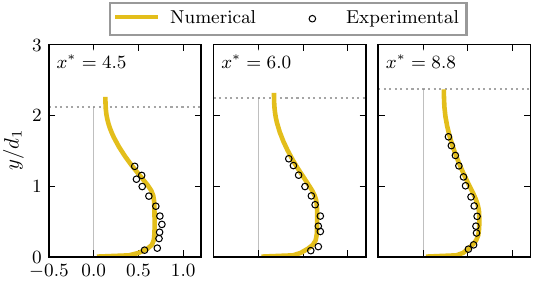}
    \caption{}\label{fig:2004_Liu_Fr2_V}
\end{subfigure}

\begin{subfigure}[b]{\columnwidth}
    \centering
    \includegraphics[width=1\columnwidth]{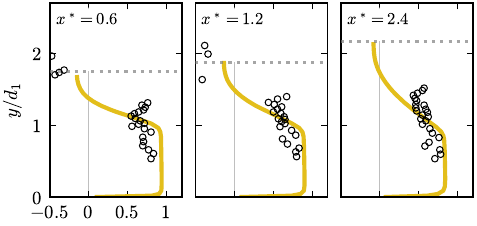}
    \caption{}\label{fig:2022_Wutrich_Fr2.4_V}
\end{subfigure}

\begin{subfigure}[b]{\columnwidth}
    \centering
    \includegraphics[width=1\columnwidth]{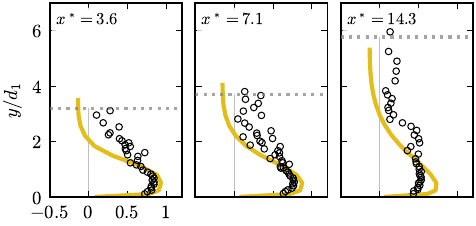}
    \caption{}\label{fig:2000_Chanson_Fr6.33_V}
\end{subfigure}

\begin{subfigure}[b]{\columnwidth}
    \centering
    \includegraphics[width=1\columnwidth]{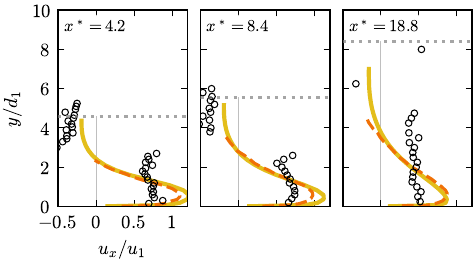}
    \caption{}\label{fig:2014_Wang_Fr7.5_V}
\end{subfigure}

\caption{Simulated (lines) and experimental (open symbols) non-dimensional streamwise velocity profiles ($u_x/u_1$) at three downstream positions. (a) Fr$_1 = 2$, (b) Fr$_1 = 2.4$, (c) Fr$_1 = 6.33$, (d) Fr$_1 = 7.5$ (dashed orange lines indicate results from Jesudhas et al.~\cite{jesudhas_modelling_2020}). The dashed horizontal line in each panel indicates the experimental $y_{90}$ position.}
\label{fig:all_velocity_profiles}
\end{figure}

Comparisons of velocity profiles at selected cross sections are presented in Figure~\ref{fig:all_velocity_profiles}. Time-averaged profiles (solid lines) are compared against experimental measurements (symbols), with the vertical dashed line at $u_x/u_1 = 0$ delineating the boundary between streamwise and reverse flow. The dashed horizontal line in each panel indicates the experimental $y_{90}$ position, providing a reference for the free-surface elevation.

For the weak jump at Fr$_1 = 2$ (Figure \ref{fig:2004_Liu_Fr2_V}), the model achieves the best agreement across all tested Froude numbers, accurately capturing the velocity field structure with MAE$~=~0.04$. The wall-jet structure is accurately reproduced at all three measurement positions ($x^* = 4.5$, $6.0$, $8.8$), and, above the wall jet, velocities transition smoothly to near-zero values, with no reverse flow detected.

At Fr$_1 = 2.4$ (Figure \ref{fig:2022_Wutrich_Fr2.4_V}), the jump transitions from undular toward oscillating behavior with increasing temporal variability. The model captures the wall-jet structure and predicts reverse flow in the upper region, though the magnitude is slightly underestimated. Agreement improves progressively with downstream distance, reaching good agreement at $x^* = 2.4$. The MAE of $0.27$ reflects these near-toe discrepancies.

For a stronger jump at Fr$_1 = 6.33$ (Figure \ref{fig:2000_Chanson_Fr6.33_V}), the wall jet displays a marked deceleration over the second and third measurement positions. In this case, a discrepancy exists between model predictions and experimental measurements in the roller region. The numerical simulations predict negative velocities indicative of recirculation, whereas the experimental data from \textcite{chanson_experimental_2000} show predominantly positive velocities or near-zero values. In the experimental work, \textcite{chanson_experimental_2000} noted substantial measurement challenges in this region, reporting that ``the data scatter is large'' in the recirculation zone due to low cross-correlation between probe tips in the highly unsteady flow. Such measurement difficulties are well-documented for velocity measurements in aerated recirculation regions~\cite{burgler_uncertainties_2024}.

The Fr$_1 = 7.5$ case (Figure \ref{fig:2014_Wang_Fr7.5_V}) also provides a valuable comparison against the IDDES of Jesudhas et al.~\cite{jesudhas_modelling_2020}. Our URANS approach employs $\approx18,500$ cells, while the 3D IDDES simulation used $\approx8$ million cells, more than $400$-fold increase. While both methodologies overestimate peak velocities, present simulations show larger deviations in the near-toe positions, with an average error of 0.33. Visual comparison reveals both numerical approaches capture the fundamental flow structure, including the wall jet profile and the transition to the roller region. Our simulations achieve comparable accuracy in the transition zone, while IDDES results do not extend into the roller itself, limiting comparison in that region. Our results underestimate the experimental velocities in the roller. Despite the substantial difference in computational cost, URANS results achieve comparable accuracy to IDDES counterparts for time-averaged velocities, making it well-suited for parametric studies where multiple simulations are required. Additional velocity validation results for Fr$_1 = 2.1$, $3.32$, $5.1$, and $8.48$ are provided in Appendix~\ref{app:velocity}.

The validation across the complete Froude number range reveals consistent model behavior for all the cases. The largest discrepancies manifest in the highly aerated roller region, where both experimental measurements and URANS modeling face significant challenges. Experimental uncertainties arise from the highly unsteady nature of recirculating flows in aerated regions~\cite{burgler_uncertainties_2024}, while RANS modeling limitations include time-averaging effects that eliminate some unsteady structures, and incomplete capturing of complex three-dimensional flow features~\cite{viti_numerical_2018}. Nevertheless, the model consistently and acceptably reproduces the characteristic hydraulic jump velocity structure --- a wall-bounded jet transitioning to recirculating flow above --- across the entire investigated Froude number range, aligning with theoretical understanding and previous experimental observations~\cite{chanson_air_1995}. For the intended engineering applications of estimating energy dissipation, predicting sequent depth ratios, and assessing stilling basin performance, the velocity field accuracy achieved represents a significant advance in computational efficiency while maintaining fidelity comparable to orders-of-magnitude more expensive hybrid RANS-LES approaches.

\subsection{Air Concentration}\label{subsec:airConcentration}
\begin{figure}[htb]
\centering
\captionsetup[subfigure]{aboveskip=0pt, belowskip=0pt}

\begin{subfigure}[b]{\columnwidth}
    \centering
    \includegraphics[width=1\columnwidth]{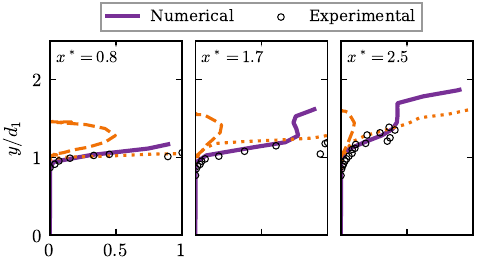}
    \caption{}\label{fig:2005_Murzyn_Fr1.98_C}
\end{subfigure}

\begin{subfigure}[b]{\columnwidth}
    \centering
    \includegraphics[width=1\columnwidth]{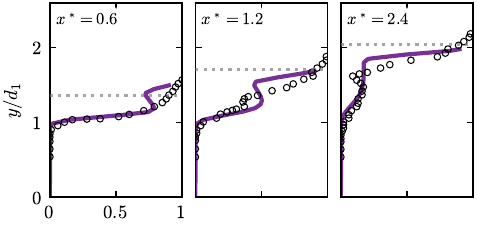}
    \caption{}\label{fig:2022_Wutrich_Fr2.1_C}
\end{subfigure}

\begin{subfigure}[b]{\columnwidth}
    \centering
    \includegraphics[width=1\columnwidth]{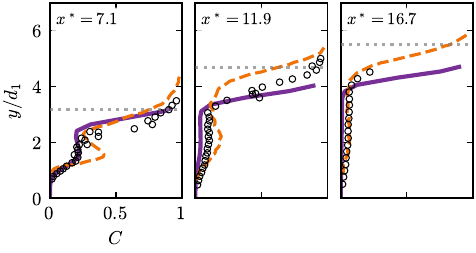}
    \caption{}\label{fig:2005_Murzyn_Fr4.8_C}
\end{subfigure}

\caption{Simulated (lines) and experimental (open symbols) air concentration profiles at low to moderate Froude numbers: 
(a) Fr$_1 = 1.98$ (dashed orange line: Numerical results from Ma et al.~\cite{ma_modeling_2011} using RANS; dotted orange line: Numerical results from Ma et al.~\cite{ma_modeling_2011} using DES), 
(b) Fr$_1 = 2.1$, 
(c) Fr$_1 = 4.8$ (dashed orange line: numerical RANS results from Witt et al.~\cite{witt_simulating_2015}, dotted orange line: numerical DES results from Witt et al.~\cite{witt_simulating_2015}). The dashed horizontal line in each panel indicates the experimental $y_{90}$ position.}
\label{fig:concentration_profiles_comparison}
\end{figure}

\begin{figure}[htb]
\centering
\captionsetup[subfigure]{aboveskip=0pt, belowskip=0pt}

\begin{subfigure}[b]{\columnwidth}
    \centering
    \includegraphics[width=1\columnwidth]{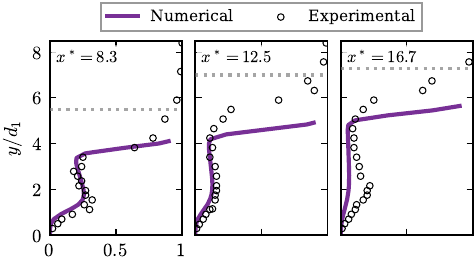}
    \caption{}\label{fig:2007_Kucukali_Fr6.9_C}
\end{subfigure}

\begin{subfigure}[b]{\columnwidth}
    \centering
    \includegraphics[width=1\columnwidth]{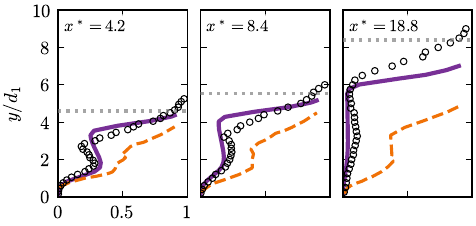}
    \caption{}\label{fig:2014_Wang_Fr7.5_C}
\end{subfigure}

\begin{subfigure}[b]{\columnwidth}
    \centering
    \includegraphics[width=1\columnwidth]{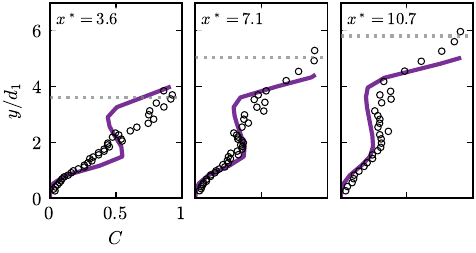}
    \caption{}\label{fig:2000_CB_Fr8.48_C}
\end{subfigure}

\caption{Simulated (lines) and experimental (open symbols) air concentration profiles at:
(a) Fr$_1 = 6.9$, 
(b) Fr$_1 = 7.5$ (dashed line indicate IDDES results from Jesudhas et al.~\cite{jesudhas_modelling_2020}), 
(c) Fr$_1 = 8.48$. The dashed horizontal line in each panel indicates the experimental $y_{90}$ position.}
\label{fig:concentration_profiles_additional}
\end{figure}

Six representative cases, spanning Fr$_1 = 1.98-8.48$, are presented in Figures \ref{fig:concentration_profiles_comparison} and \ref{fig:concentration_profiles_additional}. These cases were selected to cover the full spectrum of Fr$_1$, while prioritizing those with available numerical results from other authors to enrich the validation. Additional cases (Fr$_1 =$ 2.1, 3.32 and 6.33) are provided in Appendix~\ref{app:concentration}. All profiles show time-averaged predictions (solid lines) compared against experimental measurements (symbols). The dashed horizontal line in each panel indicates the experimental $y_{90}$ position, providing a reference for the free-surface elevation. In Table~\ref{tab:turbulence_closure_performance}, we provide quantitative assessment through MAE averaged across all measurement positions for each case.

For the Fr$_1 = 1.98$ case (Figure~\ref{fig:2005_Murzyn_Fr1.98_C}), experimental measurements from Murzyn et al.~\cite{murzyn_optical_2005} exhibit some scatter and only a few measured points. The present model captures the near-zero concentrations below the shear layer ($y/d_1 \lesssim 1$) and a gradual concentration rise through the turbulent mixing region toward the free surface. At the most downstream position ($x^* = 2.5$), the model reproduces the observed concentration profile well within the measured range, though the scarcity of data prevents definitive quantification of accuracy at the free surface. In comparison, the two-fluid RANS formulation with sub-grid air entrainment functions of Ma et al.\cite{ma_modeling_2011} (orange dashed line Figure~\ref{fig:2005_Murzyn_Fr1.98_C}) systematically underpredicts air concentrations at all three stations, with particularly pronounced underprediction at the downstream position where their model returns near-zero concentrations. Ma et al. demonstrated that their model, coupled with a DES closure ($724,000$ elements), successfully reproduced void fractions in the upper roller region (orange dotted line Figure~\ref{fig:2005_Murzyn_Fr1.98_C}), albeit at substantially higher computational cost due to the requirement of 3D simulations to resolve large-scale turbulent structures.

At Fr$_1 = 2.1$ (Figure \ref{fig:2022_Wutrich_Fr2.1_C}), air entrainment intensifies moderately with experimental concentrations reaching approximately $50\%$ in the turbulent shear layer region for the intermediate cross section. The characteristic S-shaped profile, with low concentrations in the lower part of the flow and increasing concentrations toward the free surface, is accurately reproduced at all three stations ($x^* = 0.6$, $1.2$, $2.4$), and error metrics correspond to the best agreement across all cases studied (MAE$~=~0.07$).

The Fr$_1 = 4.8$ case (Figure~\ref{fig:2005_Murzyn_Fr4.8_C}) provides a valuable comparison against bubble-resolved simulations by Witt et al.~\cite{witt_simulating_2015}, who employed meshes with $8$ cells per bubble diameter. Our approach achieves similar overall agreement to bubble-resolved results, with their downstream predictions closer to experimental points, but exhibiting overprediction of peak concentrations at upstream positions. The computational cost comparison reveals the practical advantage of our approach: Witt et al.~\cite{witt_simulating_2015} required $\approx 370,000$ cells and $289$ CPU-hours (for 15~s of simulated time), while the present approach achieves comparable accuracy with $\approx 13,000$ cells and approximately 1~CPU-hour for an equivalent simulation period --- representing $\approx 30$-fold fewer cells and $\approx 300$-fold reduced computational cost.

At Fr$_1 = 6.90$ (Figure \ref{fig:2007_Kucukali_Fr6.9_C}), the simulations reproduce the profile shape and concentration gradients within the bubbly region ($y/d_1 \lesssim 4$) reasonably well at all stations, with MAE$~=~0.09$. The primary discrepancy is a consistent underprediction of the free-surface elevation $y_{90}$, where the numerical profile finishes approximately $1$--$2~d_1$ below the experimental measurements.
The Fr$_1 = 7.5$ case (Figure \ref{fig:2014_Wang_Fr7.5_C}), presents another comparison against Jesudhas et al.~\cite{jesudhas_modelling_2020} IDDES simulations employing $\approx 8$ million cells as opposed to our $\approx 18,500$ cells. Jesudhas et al.'s results capture the characteristic S-shaped concentration profile, but overpredicts peak concentrations and underpredicts values at free-surface levels. In contrast, our URANS simulations achieve superior overall agreement at all three positions, with MAE$~=~0.09$.
At the highest investigated Froude number (Fr$_1 = 8.48$, Figure \ref{fig:2000_CB_Fr8.48_C}), the model demonstrates good performance across all three cross sections ($x^* = 3.6$, $7.1$, $10.7$), with MAE$~=~0.10$.
The systematic validation confirms that the model successfully reproduces the characteristic four-zone concentration structure identified by~\textcite{chanson_air_1995}: Near-zero concentration below the shear layer, pronounced concentration peaks within the turbulent mixing region, concentration decrease in the upper shear boundary, and subsequent increase toward the free surface. MAE remains below $0.15$ across all ten validated cases, with the lowest errors at transitional Froude numbers. A consistent underprediction of $y_{90}$ was identified at intermediate-to-high Froude numbers (Fr$_1 \approx 5$--$7$), as a consequence of underprediction of air entrainment levels in these cases. The mild increase in MAE with $Fr_1$ is consistent with the growing complexity of the air-water flow as jump intensity increases; among the possible contributing factors, the monodisperse bubble diameter assumption may become less representative at higher $Fr_1$, where increasing turbulence drives more intense shear-induced breakup and wider bubble size distributions are reported~\cite{witt_simulating_2015}.

\subsection{Jump Toe Oscillations}\label{subSec:jumpToe}
\begin{figure}[htb]
    \centering
    \includegraphics[width=\columnwidth]{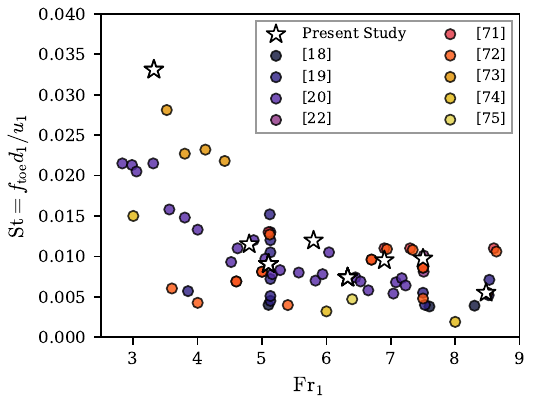}
    \caption{Strouhal numbers for jump toe oscillations as a function of Froude number. Current numerical simulations (white stars) compared against published experimental and numerical data from multiple sources~\cite{mossa_flow_1998, gualtieri_experimental_2007, chanson_convective_2010, murzyn_free_2007, chachereau_free-surface_2011, zhang_turbulence_2013, wang_turbulence_2014, jesudhas_modelling_2020}.}
    \label{fig:FrSt_toe}
\end{figure}

The jump toe exhibits quasi-periodic oscillations in the streamwise direction with characteristic frequencies that scale with jump incident Fr$_1$, driven by the cyclic formation, growth, and ejection of large-scale vortical structures within the roller region~\cite{wang_air_2015}. These oscillations couple with free-surface fluctuations and the air entrainment process, making their accurate prediction essential for dynamic load assessment on stilling basins. We present Strouhal number ($St = f_{toe} d_1/u_1$, where $f_{toe}$ is the dominant toe oscillation frequency) as a function of incident Froude number in Figure~\ref{fig:FrSt_toe}, comparing our numerical predictions (white stars) against experimental datasets from the literature~\cite{gualtieri_experimental_2007, chanson_convective_2010, murzyn_free_2007, chachereau_free-surface_2011, zhang_turbulence_2013, richard_classical_2013, wang_turbulence_2014, jesudhas_modelling_2020} across the Fr$_1 = 2-9$ range.
Time series analysis of jump toe position were conducted over 30-second periods with 10 Hz sampling frequency to capture the dominant oscillatory modes. Spectral analysis via Fast Fourier Transform (FFT) identified a single dominant frequency peak for each case, corresponding to the large-scale roller pulsation cycle.
The numerical predictions demonstrate excellent agreement with experimental observations across the entire investigated Froude number range (Figure \ref{fig:FrSt_toe}), with results consistently falling within the experimental scatter. The data reveal a clear inverse relationship between Strouhal number and Froude number: $St$ decreases from $\approx 0.05-0.06$ at Fr$_1 = 1.98$ to $0.005-0.007$ at Fr$_1 = 8.48$, consistent with previous observations~\cite{wang_air_2015}. In dimensional terms, frequencies decrease from approximately 1.5 Hz at Fr$_1 = 2$ to 1--1.2 Hz at Fr$_1 > 6$, even as approach velocities increase from 1.7 to 3.3 m/s. This physical behavior reflects the formation of progressively larger-scale vortical structures in stronger jumps, which require longer time scales to complete their formation-growth-ejection cycles~\cite{wang_air_2015}.
\begin{figure}[htb]
\centering
\captionsetup[subfigure]{aboveskip=0pt, belowskip=0pt}

\begin{subfigure}[b]{\columnwidth}
    \centering
    \includegraphics[width=1\columnwidth]{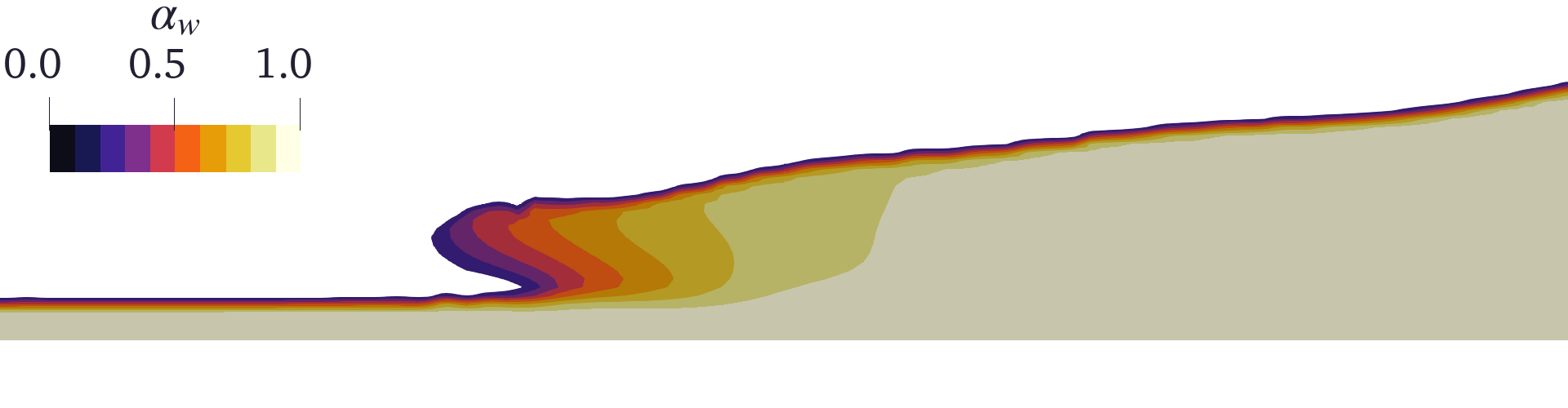}
    \caption{}\label{fig:2014_Wang_Fr7.5_11.9}
\end{subfigure}

\begin{subfigure}[b]{\columnwidth}
    \centering
    \includegraphics[width=1\columnwidth]{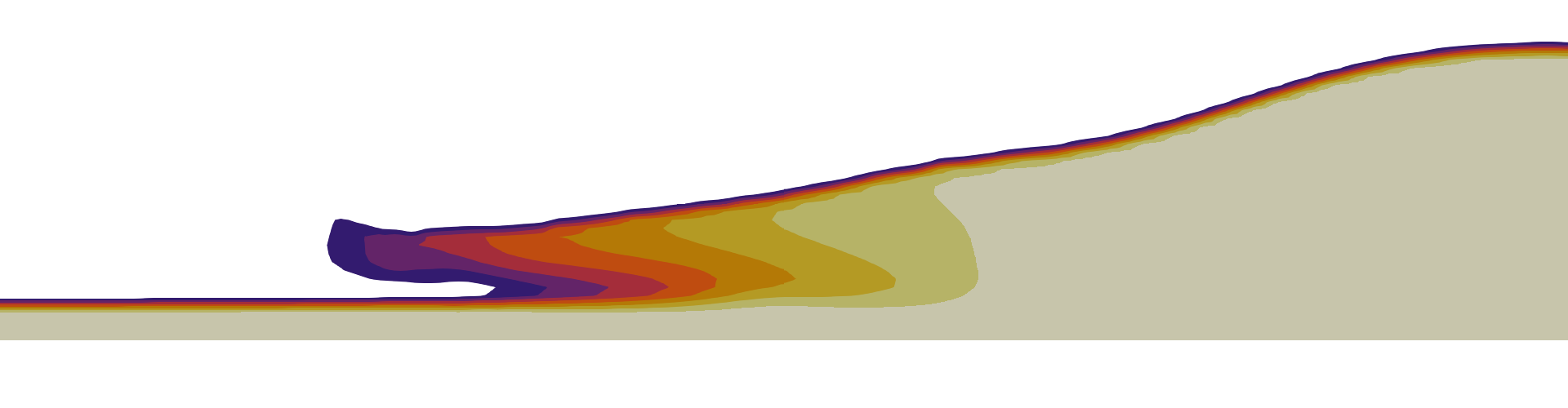}
    \caption{}\label{fig:2014_Wang_Fr7.5_11.2}
\end{subfigure}

\begin{subfigure}[b]{\columnwidth}
    \centering
    \includegraphics[width=1\columnwidth]{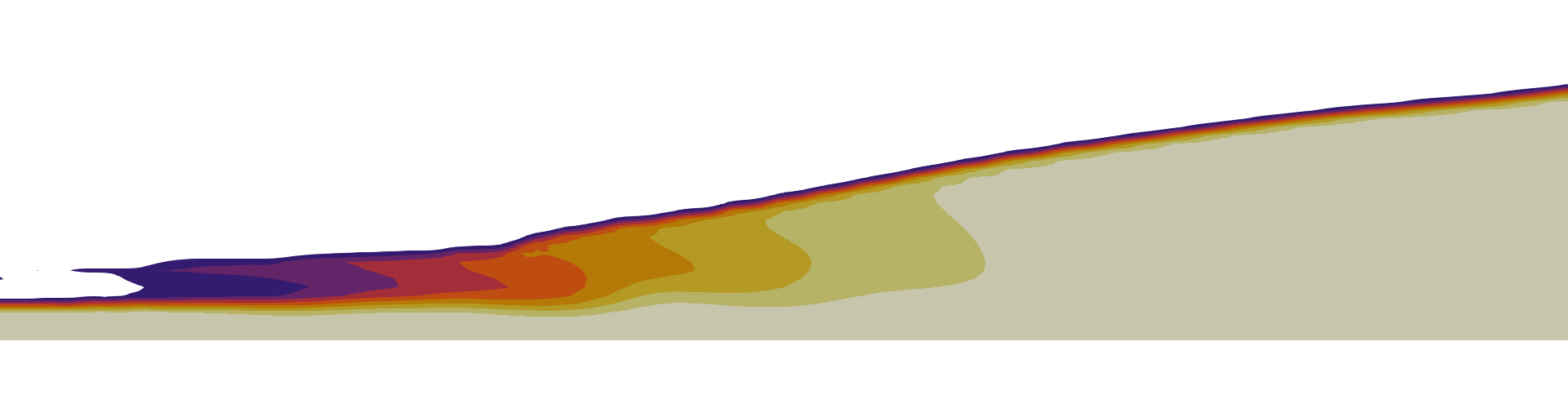}
    \caption{}\label{fig:2014_Wang_Fr7.5_11.5}
\end{subfigure}
\caption{Temporal evolution of oscillatory jump toe and roller dynamics for Fr$_1 = 7.5$. Contours show water fraction ($\alpha_w$) at three times: (a) $t = 13.8$ s, (b) $t = 14.1$ s, (c) $t = 14.4$ s. Flow is from left to right.}
\label{fig:timeSequence}
\end{figure}

The oscillation mechanism of quasi-periodic formation, growth and ejection of large-scale vortical structures is illustrated in the time sequence of Figure~\ref{fig:timeSequence}. This cyclic behavior has been extensively documented through experimental observations, where the longitudinal jump toe oscillation is coupled with the production and downstream advection of large-scale eddies in the developing shear layer~\cite{zhang_turbulence_2013, wang_air_2015}. An upstream toe position is shown in Figure~\ref{fig:2014_Wang_Fr7.5_11.5}, where roller expansion correlates with decreased surface elevation and intensified air entrainment at the impingement zone. An intermediate phase with maximum roller development and air concentration penetration into the recirculation zone is shown in Figure~\ref{fig:2014_Wang_Fr7.5_11.9}. The downstream toe movement corresponding to roller compression is depicted in Figure~\ref{fig:2014_Wang_Fr7.5_11.2}; surface elevation increases as vortical structures are ejected downstream. These oscillation amplitudes typically range from $2-10$ d$_1$ in streamwise extent.

\subsection{Location of the Free Surface}
\subsubsection*{Mean Profile}
\begin{figure}[htb]
    \includegraphics[width=\columnwidth]{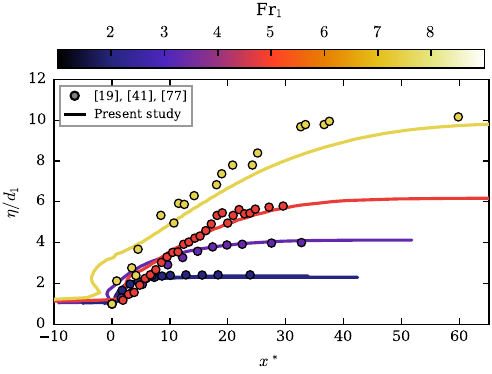}
     \caption{\label{fig:FS} Comparison of mean non-dimensional free-surface profiles ($\eta/d_1$) as a function of the non-dimensional streamwise position ($x^*$) for different inflow Froude numbers. Numerical results are shown as continuous lines; experimental data from ~\textcite{wang_turbulence_2014}, Liu et al.~\cite{liu_evaluation_2004}, and Murzyn et al.~\cite{murzyn_optical_2005} are included as filled circles.}
\end{figure}

Analysis of mean free-surface streamwise profiles ($\eta$) defines the model's ability to predict a variable of key importance in hydraulic engineering applications. Figure~\ref{fig:FS} presents non-dimensional free-surface profiles for cases spanning the investigated Froude number range, where complete experimental profiles were available in the literature~\cite{liu_evaluation_2004, murzyn_optical_2005, wang_turbulence_2014}.
Agreement is very good to excellent for weak to moderate jumps (Fr$_1 \leq 3.32$), where predicted profiles closely match experimental measurements throughout the domain. For Fr$_1 = 5.1$, the model captures the overall profile shape, though free-surface elevations are slightly overpredicted near the toe. For the strongest jump (Fr$_1 = 7.5$), numerical predictions fall below the experimental measurements for $x^* > 10$.

%To further assess model performance in the low-Froude regime, we include recent experimental data from Hu and Chanson~\cite{HuChanson2024}, who conducted detailed free surface measurements in hydraulic jumps at Fr$_1 = 2.1$, $2.6$, $3.1$, and $3.8$ with Reynolds numbers between $1.3 - 2.5 \times 10^5$. Although the specific flow conditions ($d_1$, $u_1$, channel geometry) differ from our simulations, their data (Figure~\ref{fig:FS}, squares) exhibit consistent agreement with both our numerical predictions and the previously validated experimental datasets in the overlapping Froude number range.

\subsubsection*{Oscillations of the Free Surface}
\begin{figure}[htb]
    \includegraphics[width=1\columnwidth]{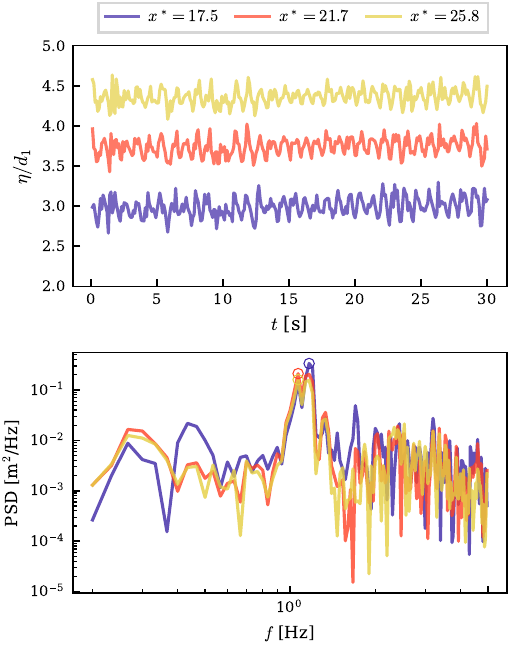}
    \caption{\label{fig:fsFrequency_Fr5.1} Temporal evolution of maximum free-surface elevation $\eta/d_1$ (top) and corresponding power spectrum density (bottom) at three longitudinal positions within the roller region for Fr$_1 = 5.1$.}
\end{figure}

\begin{figure}[htb]
    \includegraphics[width=1\columnwidth]{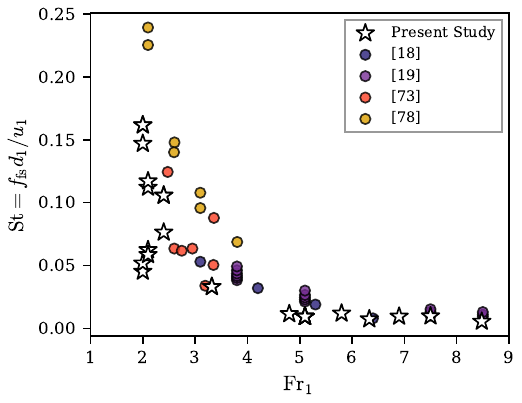}
    \caption{\label{fig:FrSt_freeSurface} Strouhal number for free-surface oscillations as a function of incident Froude number. Current numerical simulations (white stars) compared against published experimental data from multiple sources~\cite{murzyn_free-surface_2009, chachereau_free-surface_2011, wang_turbulence_2014, HuChanson2024}.}
\end{figure}

In Figure~\ref{fig:fsFrequency_Fr5.1}, we include representative time series and spectral analysis of the position of the free surface at given cross sections (Fr$_1 = 5.1$), showing a dominant oscillation frequency between 1.07--1.17 Hz across three longitudinal positions within the roller. Our numerical predictions are obtained using spectral analysis via Fast Fourier Transform of surface elevation time series extracted at multiple fixed spatial locations within the roller region over 30-second post-transient periods.
In Figure~\ref{fig:FrSt_freeSurface}, we present characteristic oscillation frequencies expressed as Strouhal numbers ($St = f_{fs}$ $d_1/u_1$, where $f_{fs}$ is the dominant free-surface frequency) across Froude numbers $1.98-8.48$, comparing our numerical predictions against experimental databases from~\textcite{murzyn_free_2007} and~\textcite{wang_turbulence_2014}. These authors conducted extensive analysis of frequency modes across multiple longitudinal positions within the roller.

Experimental spectra if the free-surface locations~\cite{wang_turbulence_2014, wang_air_2015} reveal two distinct frequencies with different physical origins. Primary high-frequency oscillations ($f_1$), with Strouhal numbers ranging from $St \approx 0.015-0.055$ depending on Froude number, associate with high-frequency vortex shedding and small-scale turbulent structures. Secondary low-frequency oscillations ($f_2$), with Strouhal numbers ranging from $St \approx 0.003-0.015$, linked to large-scale jump oscillations, representing the dominant mode of roller pulsation that governs the quasi-periodic behavior observable in hydraulic jumps.
Our URANS simulations successfully capture the secondary low-frequency mode ($f_2$) across the moderate to strong jump range (Fr$_1 \geq 5$), where experimental data for this mode are available. The white stars in Figure~\ref{fig:FrSt_freeSurface} fall consistently near the $f_2$ data in this range, validating the model's capability to reproduce the dominant large-scale instability mechanism identified by~\textcite{wang_turbulence_2014}. For weaker jumps (Fr$_1 < 5$), experimental secondary frequencies are not reported in the available datasets, while the primary high-frequency mode ($f_1$) data fall, as expected, above our model predictions.
The correspondence between predicted toe oscillation frequencies (Figure~\ref{fig:FrSt_toe}) and free-surface secondary frequencies (Figure~\ref{fig:FrSt_freeSurface}) at Fr$_1 \geq 5$ provides strong evidence for the coupled nature of hydraulic jump dynamics at these conditions, consistent with the experimental observations of~\textcite{wang_air_2015}. Both phenomena exhibit similar Strouhal number dependencies on Froude number, with predictions falling along the same trend as the large-scale oscillatory mode ($f_2$). This correspondence demonstrates that toe movement, roller pulsation, and surface elevation fluctuations represent different observable manifestations of the same underlying large-scale mechanism that dominates the quasi-periodic behavior of moderate to strong hydraulic jumps.

%\subsubsection{Limitations of the URANS Approach for Hydraulic Jump Simulations}
Whereas the present URANS simulations successfully capture the dominant large-scale oscillatory phenomena (toe movement, roller pulsation, and associated free-surface fluctuations), certain limitations inherent to the modeling approach should be acknowledged. The URANS framework resolves the mean flow and large-scale unsteady motions, while modeling all turbulent fluctuations through eddy viscosity closures. Consequently, high-frequency small-scale turbulent structures are not resolved. RANS and URANS approaches are well-suited for applications where time-averaged quantities are of primary interest, such as mean velocity and pressure distributions for structural loading assessment, time-averaged air concentration profiles for dissolved oxygen predictions, and roller geometry and sequent depth estimation for stilling basin design. However, they are less appropriate for analyses requiring instantaneous flow characteristics, such as cavitation risk assessment, structural vibration and peak loads calculations, and studies where accurate turbulence assessment is critical.

%% file: VI_Conclusions.tex
\section{Conclusions}\label{sec:Conclusions}

This study presented the first application of a three-phase mixture model to hydraulic jumps across an unprecedented wide range of Froude numbers (Fr$_1 = 1.98$--$8.48$). To the best of the authors' knowledge, this is the first comprehensive assessment examining velocity fields, air concentration distributions, toe oscillation dynamics, and free-surface evolution spanning undular to strong jumps using a URANS approach.

Three main findings emerge from this investigation:
\begin{enumerate}
    \item The model achieves good agreement with experimental data for both velocity and air concentration fields across the investigated Froude number range, with MAE ranging from 0.04 (Fr$_1$ = 2) to 0.33 (Fr$_1$ = 7.5), and accuracy comparable to IDDES simulations at Fr$_1 = 7.5$~\cite{jesudhas_modelling_2020} 
    despite using approximately $400\times$ fewer cells. Air concentration predictions reproduce the characteristic four-zone structure~\cite{chanson_air_1995} across all conditions, with MAE $\leq 0.15$ in all ten validated cases. At Fr$_1 =$ 4.8, comparable accuracy to bubble-resolved simulations~\cite{witt_simulating_2015} is achieved with $\approx 30\times$ fewer cells and $\approx 300\times$ reduced computational cost. An underprediction of $y_{90}$ at intermediate-to-high Froude numbers (Fr$_1 \approx 5$--$7$), and underestimation of reverse flow velocities in highly aerated roller regions, represent the primary limitations of the present approach.
    \item The model successfully reproduces quasi-periodic large-scale oscillations. Predicted toe oscillation Strouhal numbers fall consistently within experimental scatter for Fr$_1 = 2-9$, with frequencies decreasing from $\approx 1.5$ Hz in weak jumps to $\approx 1-1.2$ Hz in strong jumps. The correspondence between toe and free-surface oscillation frequencies confirms their coupled nature as manifestations of the same underlying large-scale instability mechanism.
    \item Turbulence closure selection influences air entrainment prediction through a density-mediated mechanism. Intense air entrainment at the jump toe reduces local mixture density from $\rho_m \approx$ 1000 to 250–500 kg/m$^3$, affecting the turbulent kinetic energy field differently, depending on the eddy viscosity formulation. We suggest that the $k$-$\varepsilon$ family artificially overestimates $\nu_t$ through a self-reinforcing feedback mechanism, and the resulting excessive bubble diffusion must be controlled through a Schmidt number increase. In contrast, the $k$-$\omega$ SST closure suppresses this spurious production mechanism before it develops, rendering the model naturally insensitive to the Schmidt number. Selecting $k$-$\omega$ SST with $Sc_t = 1$ therefore avoids empirical parameter adjustment while achieving accuracy comparable to RNG $k$-$\varepsilon$ ($Sc_t = 3$) across the full Froude number range.
\end{enumerate}

The approach provides engineering-sound accuracy for stilling basin design, spillway optimization, and energy dissipation assessment.

Future research directions include: (1) incorporation of polydisperse bubble population formulations to better represent bubble size distributions and breakup/coalescence processes; (2) development or adaptation of turbulence models specifically designed for intense two-phase flows with strong density variations; (3) extension to fully three-dimensional configurations to capture transverse flow features and sidewall effects; and (4) application to other hydraulic jump configurations including submerged jumps and jumps with energy dissipation devices.

%% file: VII_Appendix.tex
\appendix
\section{Statistical Stationarity Assessment}\label{app:stationarity}
\begin{figure}[htb]
\centering
\includegraphics{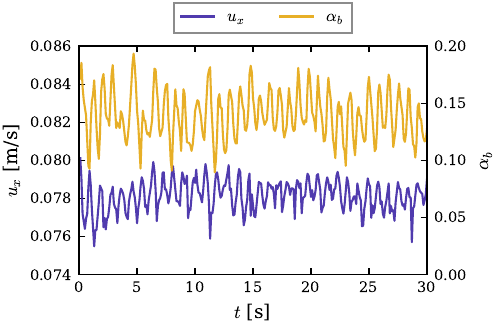}
\caption{\label{fig:2007_Murzyn_Fr5.1_stationarity}Temporal evolution of volume-integrated streamwise velocity component ($u_x$, left axis) and bubble fraction ($\alpha_b$, right axis) for the Fr$_1 =$ 5.1 hydraulic jump simulation.}
\end{figure}

Assessment of statistical stationarity is essential for obtaining trustful time-averaged results from unsteady simulations. In Figure~\ref{fig:2007_Murzyn_Fr5.1_stationarity} the temporal evolution of two volume-integrated quantities for the Fr$_1 = 5.1$ case are presented: The bubble phase fraction ($\alpha_b$) and the streamwise velocity component ($u_x$). 
Both quantities exhibit sustained quasi-periodic oscillations characteristic of the large-scale dynamics of hydraulic jumps throughout the 30-second post-transient averaging period.

%%%%%%%%%%%%%%%%%%%%%%%%%%%%%%%%%%%%%%%%%%%%%%%%%%%%%%%%%%%%%%%%

\section{Mesh-Independence Tests and Grid-Convergence Analysis}\label{app:Mesh}
Mesh-independence tests and grid convergence analysesfollowing the methodology of Celik et al.~\cite{celik_procedure_2008} were developed. Three grid sizes were employed: coarse (7 mm), medium (5 mm), and fine (3.5 mm), with refinement ratio of $r \approx \sqrt{2}$. The analysis focused on roller length, spatially averaged peak streamwise velocity, and spatially averaged peak air concentration in the shear layer for the case with Fr$_1 = 8.48$.

\begin{figure}[htb]
\centering
\captionsetup[subfigure]{aboveskip=0pt, belowskip=0pt}

\begin{subfigure}[b]{\columnwidth}
    \centering
    \includegraphics[width=1\columnwidth]{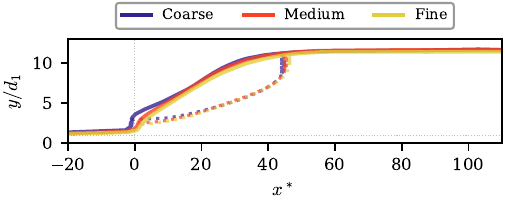}
    \caption{}\label{fig:mesh_conv_fs+roller}
\end{subfigure}

\begin{subfigure}[b]{\columnwidth}
    \centering
    \includegraphics[width=1\columnwidth]{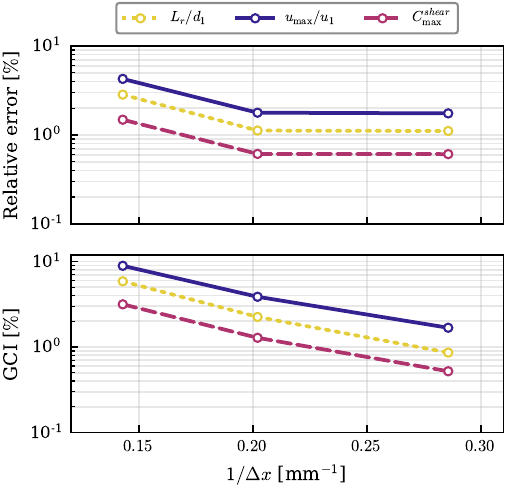}
    \caption{}\label{fig:mesh_conv_error}
\end{subfigure}

\caption{Mesh-independence and grid convergence analysis for hydraulic jump at Fr$_1 = 8.48$:
(a) free surface profiles and stagnation streamlines for three grid resolutions;
(b) relative error and Grid Convergence Index (GCI) versus $1/\Delta x$ for roller length ($L_r/d_1$), spatially averaged peak velocity ($\overline{u}_{\max}/u_1$), and spatially averaged peak air concentration in the shear layer ($\overline{C}_{\max}^{\rm shear}$).}
\label{fig:mesh_convergence}
\end{figure}

Free-surface profile and stagnation streamlines are presented in Figure \ref{fig:mesh_conv_fs+roller} for the three grid resolutions. The consistency across all three grid resolutions confirms that both the roller length and free surface elevation are well-resolved, with the medium and fine meshes showing an overlap.

The convergence of relative error and Grid Convergence Index (GCI) as functions of the inverse mesh size $1/\Delta x$ are presented in Figure \ref{fig:mesh_conv_error}. The relative error decreases monotonically with mesh refinement for all quantities. GCI 
values remain below $1.7$\% for the fine mesh across all variables, and below $5$\% for the medium mesh. Based on Figures~\ref{fig:mesh_conv_fs+roller} and \ref{fig:mesh_conv_error} and Table~\ref{tab:gci_results}, the medium mesh ($\Delta x = \Delta y = 5$ mm) was selected for all simulations.

%%%%%%%%%%%%%%%%%%%%%%%%%%%%%%%%%%%%%%%%%%%%%%%%%%%%%%%%%%%%%%%%

\section{Additional Velocity Validation Results}\label{app:velocity}
\begin{figure}[htb]
\centering
\captionsetup[subfigure]{aboveskip=0pt, belowskip=0pt}

\begin{subfigure}[b]{\columnwidth}
    \centering
    \includegraphics[width=1\columnwidth]{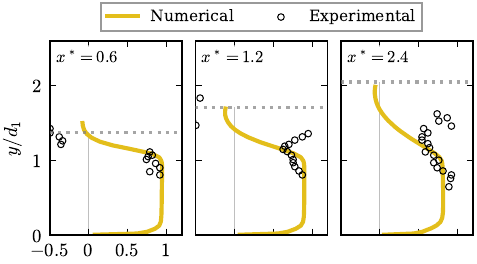}
    \caption{}\label{fig:2022_Wutrich_Fr2.1_V}
\end{subfigure}

\begin{subfigure}[b]{\columnwidth}
    \centering
    \includegraphics[width=1\columnwidth]{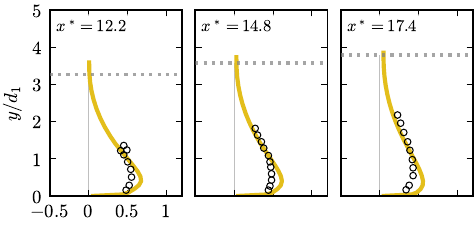}
    \caption{}\label{fig:2004_Liu_Fr3.32_V}
\end{subfigure}

\begin{subfigure}[b]{\columnwidth}
    \centering
    \includegraphics[width=1\columnwidth]{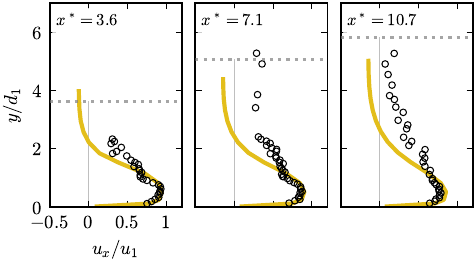}    \caption{}\label{fig:2000_CB_Fr8.48_V}
\end{subfigure}

\caption{Comparison among simulated (lines) and experimental (open symbols) non-dimensional streamwise velocity profiles ($u_x/u_1$) at different Froude numbers and downstream positions: (a) $\mathrm{Fr}_1 = 2.1$, (b) $\mathrm{Fr}_1 = 3.32$, (c) $\mathrm{Fr}_1 = 8.48$. The dashed horizontal line in each panel indicates the experimental $y_{90}$ position.}
\label{fig:velocity_appendix}
\end{figure}

This appendix presents velocity profile comparisons for cases not shown in the main text, complementing the validation presented in Section~\ref{subsec:velocity}. All profiles in Figure \ref{fig:velocity_appendix} show time-averaged numerical predictions (solid lines) compared against experimental measurements (open symbols).

%%%%%%%%%%%%%%%%%%%%%%%%%%%%%%%%%%%%%%%%%%%%%%%%%%%%%%%%%%%%%%%%
\section{Additional Air Concentration Validation Results}\label{app:concentration}

\begin{figure}[htpb]
\centering
\captionsetup[subfigure]{aboveskip=0pt, belowskip=0pt}

\begin{subfigure}[b]{\columnwidth}
    \centering
    \includegraphics[width=1\columnwidth]{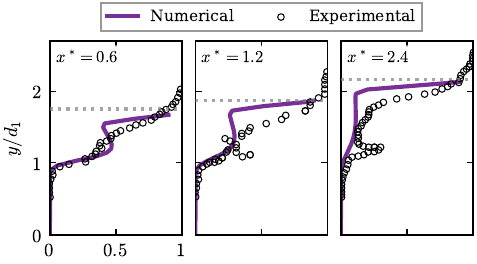}
    \caption{}\label{fig:2022_Wutrich_Fr2.4_C}
\end{subfigure}

\begin{subfigure}[b]{\columnwidth}
    \centering
    \includegraphics[width=1\columnwidth]{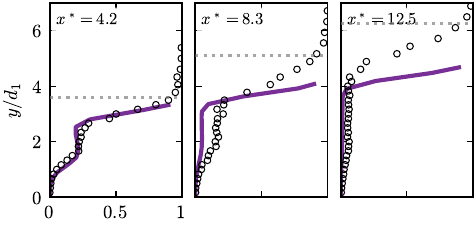}
    \caption{}\label{fig:2007_Murzyn_Fr5.1_C}
\end{subfigure}

\begin{subfigure}[b]{\columnwidth}
    \centering
    \includegraphics[width=1\columnwidth]{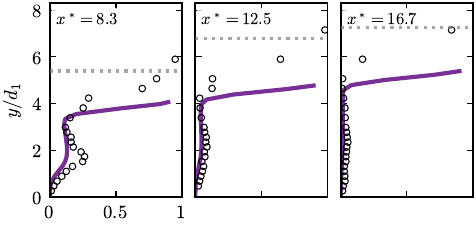}
    \caption{}\label{fig:2007_Kucukali_Fr5.8_C}
\end{subfigure}

\begin{subfigure}[b]{\columnwidth}
    \centering
    \includegraphics[width=1\columnwidth]{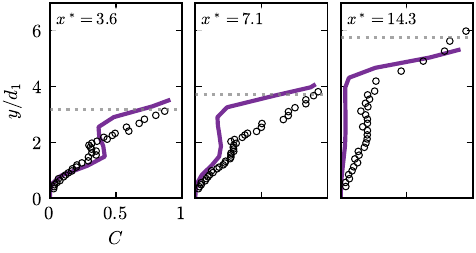}
    \caption{}\label{fig:2000_Chanson_Fr6.33_C}
\end{subfigure}

\caption{Simulated (lines) and experimental (open symbols) air-concentration profiles for additional validation cases: (a) $\mathrm{Fr}_1 = 2.4$, (b) $\mathrm{Fr}_1 = 5.1$, (c) $\mathrm{Fr}_1 = 5.8$, (d) $\mathrm{Fr}_1 = 6.33$. The dashed horizontal line in each panel indicates the experimental $y_{90}$ position.}
\label{fig:concentration_appendix}
\end{figure}

This appendix presents air concentration profile comparisons for cases not shown in the main text, complementing the validation presented in Section~\ref{subsec:airConcentration}. All profiles in Figure \ref{fig:concentration_appendix} show time-averaged numerical predictions (solid lines) compared against experimental measurements (open symbols).

